\documentclass[fleqn,usenatbib]{mnras}

\usepackage{newtxtext,newtxmath}

\usepackage[T1]{fontenc}

\DeclareRobustCommand{\VAN}[3]{#2}
\let\VANthebibliography\thebibliography
\def\thebibliography{\DeclareRobustCommand{\VAN}[3]{##3}\VANthebibliography}

\usepackage{graphicx}	
\usepackage{amsmath}	
\usepackage{xcolor}
\usepackage{CJKutf8}

\title[YSO associations from Gaia DR3]{Characterising the Kinematics and Evolution of Young Stellar Groups within 1\,kpc of the Sun Using Gaia DR3}

\author[L. F. Ding et al.]{
Long-Fei Ding,$^{1}$
Yun-Qian Li,$^{1}$
Bing-Qiu Chen,$^{1}$\thanks{E-mail: bchen@ynu.edu.cn (BQC)}
Guang-Xing Li,$^{1}$\thanks{E-mail: gxli@ynu.edu.cn (GXL)}
and Hai-Bo Yuan$^{2,3}$
\\
$^{1}$South-Western Institute for Astronomy Research, Yunnan University, Kunming, Yunnan 650091, P.\,R.\,China\\
$^{2}$Institute for Frontiers in Astronomy and Astrophysics, Beijing Normal University, Beijing 102206, P.\,R.\,China\\
$^{3}$School of Physics and Astronomy, Beijing Normal University, Beijing 100875, P.\,R.\,China\\
}

\date{Accepted XXX. Received YYY; in original form ZZZ}

\pubyear{\the\year{}}

\begin{document}
\label{firstpage}
\pagerange{\pageref{firstpage}--\pageref{lastpage}}
\maketitle

\begin{abstract}
Star-forming regions are key to understanding the formation and early evolution of stars. Young stellar objects (YSOs) form groups with distinct kinematic and spatial properties, inherited from the turbulent dynamics of their parent molecular clouds. The high-precision astrometry and photometry from Gaia Data Release 3 (DR3) enable detailed studies of these groups’ three-dimensional motions and their evolutionary stability. This study aims to investigate the kinematic properties and evolutionary consistency of YSO associations in the solar neighbourhood. Here, we show that HDBSCAN clustering of Gaia DR3 data yields 145 YSO groups comprising 5713 stars within 1 kpc, with a derived Larson’s relation of $\sigma_v = (1.10 \pm 0.13) \times r^{0.38 \pm 0.03}$, consistent across age bins up to 20 Myr. This slope aligns with the canonical value of 0.38 and typical ranges of 0.4--0.5. The stable Larson’s relation across ages indicates that the inherited turbulent structure from parent clouds persists without significant disruption. These findings establish a benchmark for studying the kinematic legacy of star-forming regions.
\end{abstract}



\section{Introduction}

Stars form from the collapse of molecular clouds, a process shaped by the interplay of gravity, turbulence, and magnetic field \citep{fleck1980, larson1981, mac2004, crut2012}. Understanding the kinematic properties of molecular clouds at different evolutionary stages provides critical insights into the mechanisms driving star formation and the dynamic evolution of the interstellar medium. The internal motions within these clouds, characterized by their velocity dispersion, reveal how energy is transferred across scales, influencing the formation of stars and the structure of galaxies. Investigating these motions helps us unravel the complex physical processes governing the lifecycle of molecular clouds and their role in shaping the Galactic environment \citep{elme2004, balle2011, izqui2021, xie2025}.

A foundational discovery in this field is Larson's relation, which describes a positive correlation between a molecular cloud's velocity dispersion, $\sigma_v$, and its size, $r$, expressed as $\sigma_v \propto r^{\beta}$, with $\beta \approx 0.38$ \citep{larson1981}. This relation suggests that velocity dispersion increases with spatial scale, a hallmark of turbulent motion where kinetic energy is injected at large scales, cascades to smaller scales, and eventually dissipates \citep{kolmo1941}. Turbulence is widely regarded as a key driver of this scaling relation, shaping the internal dynamics of molecular clouds. Larson's findings have been supported by subsequent studies, including \citet{solomon1987}, \citet{may1997}, and \citet{thomas2016}, which confirm the robustness of this scaling across various molecular cloud populations.

While the power-law index $\beta \approx 0.5$ is widely accepted, studies have reported variations in this value. Some works, such as \citet{caselli1995} and \citet{huang2023}, find smaller values of $\beta$, while others, such as \citet{zhou2022} and \citet{sun2024}, report larger values. \citet{Miville2017} constructed a catalogue of 8107 molecular clouds covering the entire Galactic plane and found that Larson's relation can be observed across a large sample of clouds, but exhibits significant scatter and may vary with cloud surface density and Galactic environment. These deviations likely reflect differences in the physical conditions of molecular clouds, such as the strength of self-gravity, magnetic fields, Galactic shear, or stellar feedback from winds and supernovae \citep{fleck1980, elme2004, ligx2022, ligx2024, xie2025}. The diversity in $\beta$ suggests that Larson's relation may encode distinct evolutionary pathways for molecular clouds, influenced by both internal and external processes. However, a comprehensive understanding of how these factors shape the scaling relation across different cloud ages remains incomplete, motivating further investigation.

Directly measuring the three-dimensional kinematic properties of molecular clouds is challenging due to the difficulty of observing gas motions in all three dimensions. Molecular clouds are often studied through line-of-sight velocity measurements \citep{may1997, jackson2006, thomas2016, sun2024}, which provide only partial information about their internal dynamics. This limitation makes it difficult to fully characterize the interplay of turbulence, gravity, and other forces within clouds. As a result, alternative approaches are needed to probe the kinematic properties of molecular clouds and their connection to star formation processes.

\begin{figure*}
	\includegraphics[width=0.9\textwidth]{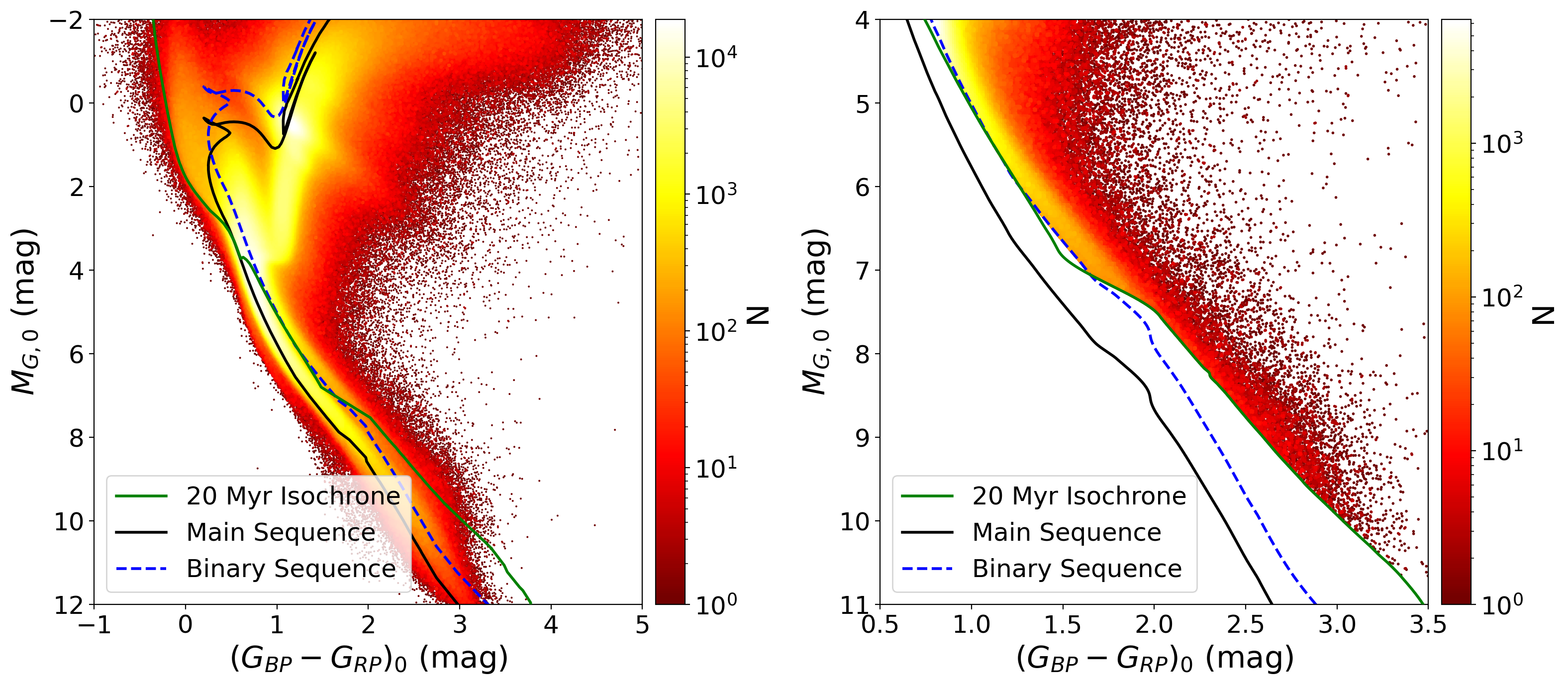}
    \caption{Left Panel: HR diagram of the sample after distance and extinction corrections. Right Panel: Enlarged view of the YSO candidates selection in the low-temperature region. The color scale indicates stellar density. The black solid line marks the main sequence, the blue dashed line represents the binary sequence, and the green solid line corresponds to a 20~Myr isochrone.}
    \label{fig:hrd}
\end{figure*}

Young stellar objects (YSOs), formed in the early stages of stellar evolution, offer a powerful means to study the kinematics of their parent molecular clouds. The evolutionary stages of YSOs are characterised by their spectral energy distributions (SEDs). Early studies classified young stellar sources based on infrared SEDs \citep{Lada1987}; this classification was subsequently extended to include the earliest embedded stages \citep{Andre1993, Greene1994}. In this framework, YSOs are typically divided into four classes, from Class~0 to Class~III \citep{Lada1987, Allen2004, Dunham2014, Marton2023, Ma2024}. Class~0/I sources represent the youngest and most deeply embedded phase, in which the central protostar is still surrounded by a dense envelope. Class~II/III sources are more evolved, with less circumstellar material, and their emission is dominated by the stellar photosphere. These stars, located near their birthplaces, retain the spatial and kinematic signatures of the gas from which they formed. Their velocities closely match those of the surrounding molecular gas, making them excellent tracers for studying Galactic structure and dynamics. For instance, \citet{zari2018} used young stellar populations, including pre-main-sequence (PMS) sources and upper-main-sequence (UMS) stars, to map structures near the Sun, while \citet{lim2021} and \citet{ha2021} explored the kinematics of Orion~A and turbulent gas motions, respectively. Similarly, \citet{li2022} and \citet{zhou2024} investigated the velocity structure of the Gould Belt Radcliffe Wave and the distribution of YSOs relative to interstellar gas. More recently, \citet{Chen2025} used the full three-dimensional velocity information of young O--B2 stars to reconstruct a superbubble structure in the Perseus arm and reveal how stellar feedback regulates molecular clouds and the star formation process. These studies highlight the utility of young stellar populations in tracing the complex physical processes of molecular clouds.

The physical origin of the observed kinematics in young stellar groups has been examined in several recent studies. \citet{Kounkel2020b} attributed the large-scale expansion of the Orion complex to a supernova explosion $\sim$6~Myr ago; \citet{Grossschedl2021} found that the gas clouds in the Orion complex were closest $\sim$6~Myr ago and are expanding radially at the 100-pc scale, suggesting a major feedback event; \citet{Drew2021} detected kinematic perturbations in the proper motions of OB stars in the far Carina Arm; \citet{Quintana2022} found that the expansion of OB stars in Cygnus is consistent with the turbulent velocity field of the primordial molecular cloud; and \citet{Grossschedl2026} studied the evolution of velocity dispersion in Sco-Cen and concluded that stellar feedback is its primary driver.

Recent advancements in observational data, particularly from the \textit{Gaia} mission, have revolutionized our ability to study YSOs and their kinematics. The \textit{Gaia} Data Release~3 (DR3) provides high-precision measurements of magnitudes, parallaxes, and proper motions for billions of stars in the Milky Way \citep{Gaia2016, gaia2023}. These data enable the identification of young stellar associations, the determination of their ages, and the analysis of their three-dimensional kinematic properties \citep{kounkel2020a, Kerr2023, Quintana2026}. By leveraging \textit{Gaia} DR3, we can systematically search for YSOs, confirm their membership in young stellar associations, and explore how kinematic properties, such as velocity dispersion, vary with association age and size.

In this study, we utilize YSOs as tracers to investigate the kinematic properties of their parent molecular clouds, focusing on the relationship between velocity dispersion and size. Using \textit{Gaia} DR3, we identify 145 young stellar associations containing 5713 member stars. We analyse their spatial distribution and, for the first time, examine how Larson's relation varies across associations of different ages. This approach allows us to probe the interplay between turbulence, gravity, and stellar feedback in shaping the evolution of molecular clouds, offering new insights into the dynamical processes governing star formation in the Milky Way.

\begin{figure*}
	\includegraphics[width=0.9\textwidth]{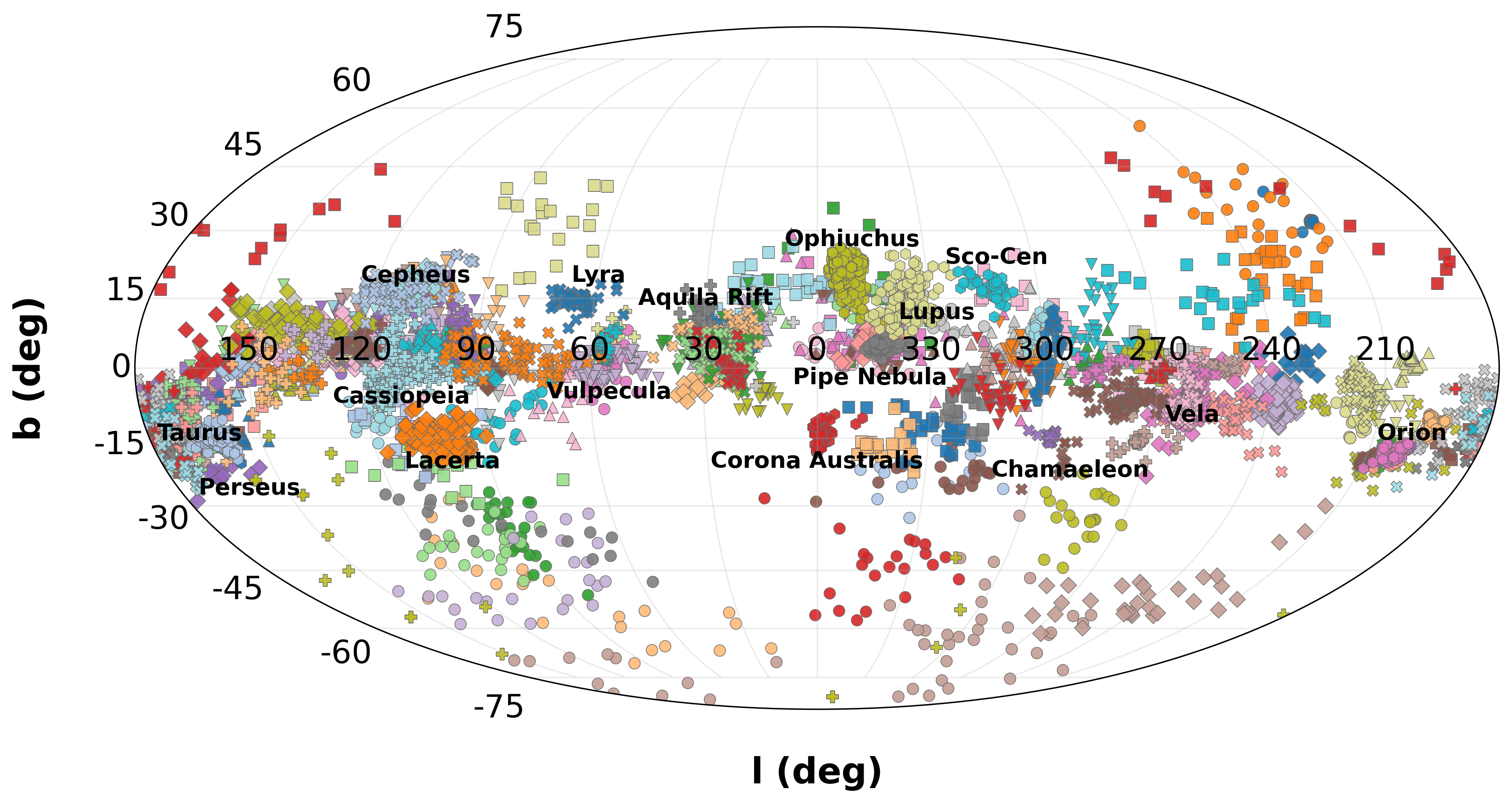}
    \caption{Distribution of YSO association member stars identified by HDBSCAN in Galactic coordinates. Points with different colours and marker styles represent different clustering groups. The black labels mark the names of molecular clouds in the solar neighbourhood \citep{zari2018, Zucker2023}.}
    \label{fig:yso_lb}
\end{figure*}

\section{Data}
This study utilizes data from \textit{Gaia} DR3 \citep{gaia2023}, which provides astrometric measurements and broadband $G$, $G_{\rm BP}$, and $G_{\rm RP}$ photometry for over one billion stars. Additionally, \textit{Gaia} DR3 includes mean radial velocities for approximately 33 million stars and low-resolution XP spectra for about 220 million stars, calibrated for absolute flux. The primary goal of this work is to identify YSOs, perform clustering analysis, and investigate the three-dimensional kinematics of YSO associations. To achieve this, we use \textit{Gaia}'s multi-band photometry, astrometric data (including parallaxes and proper motions), radial velocities, and stellar extinction estimates derived from XP spectra \citep{Zhang2025}. Since radial velocity (RV) data are critical for our analysis, we begin by constructing our sample from the 33 million stars with available RV measurements. We note that Gaia DR3 radial velocities are available only for a relatively bright subset of Gaia sources, with a typical limit of $G_{\rm RVS} \lesssim 14$~mag, where $G_{\rm RVS}$ is the Gaia Radial Velocity Spectrometer magnitude \citep{Katz2023}. Class~0/I YSOs remain deeply embedded in their protostellar envelopes and are therefore significantly more difficult to detect in optical surveys than the more evolved Class~II/III YSOs \citep{Lada1987, Dunham2014}. Consequently, embedded YSOs, particularly Class~0/I sources, are likely underrepresented in our RV-selected sample. The sample used in this work should therefore be regarded as a Gaia RV-selected, optically visible PMS/YSO candidate sample suitable for six-dimensional kinematic analysis, rather than a complete census of all YSO evolutionary stages.

To ensure data reliability, we apply quality criteria from \citet{lindegren2018}, specifically equations C.1 and C.2, to filter out sources with spurious high parallaxes and to mitigate photometric errors in the $G_{\rm BP}$ and $G_{\rm RP}$ bands, particularly for faint sources or those in crowded regions. We further require a parallax precision of at least 20 per cent, corresponding to $\varpi/\sigma_{\varpi} \geq 5$, where $\varpi$ is the parallax and $\sigma_{\varpi}$ its uncertainty, to ensure accurate distance measurements.

\section{YSO Candidate Selection and Clustering}
In this study, we first identify YSO candidates from the Hertzsprung--Russell (HR) diagram and then apply the HDBSCAN clustering algorithm to group them into associations.

\subsection{Distance and Extinction Corrections}
To construct the HR diagram, we correct the observed colours and magnitudes of our sample stars for distance and extinction. Distances are derived from \textit{Gaia} DR3 parallaxes using a Bayesian approach \citep{Ma2005, BailerJones2018, Chen2019OBStar, Shen2022, Chen2025}, with the posterior probability defined as
\begin{equation}
    p(d \mid \varpi) = d^2 \exp\left[-\frac{1}{2\sigma_{\varpi}^2}\left(\varpi - \varpi_{\rm zp} - \frac{1}{d}\right)^2\right] p(d),
\end{equation}
where $\sigma_{\varpi}$ is the parallax uncertainty, $\varpi_{\rm zp}$ is the global parallax zero-point offset, and $p(d)$ is the prior on the spatial density distribution of the sample stars. We adopt a zero-point correction of $\varpi_{\rm zp} = -0.017\,\mathrm{mas}$ \citep{gaia2023} and use the Galactic structure model from \citet{Chen2017} as the spatial density prior.

For extinction corrections, we rely on stellar extinction estimates from \citet{Zhang2025}, who derived atmospheric parameters and extinction values for 220 million stars using \textit{Gaia} XP spectra and 2MASS near-infrared photometry. Since their training sample did not specifically include YSOs, their extinction estimates for such objects may be less reliable. To address this, we estimate the extinction for each star in our sample by averaging the extinction values of its six nearest neighbours, identified using the $k$-nearest neighbours (KNN) algorithm \citep{scikit-learn}. We select stars from the \citet{Zhang2025} catalogue with $\mathrm{quality\_flags} < 8$, following their recommendations for reliable extinction. Since both our distances and those in \citet{Zhang2025} are based on \textit{Gaia} parallaxes, they are mutually consistent, allowing us to identify the nearest stars for each star in our sample. If the farthest of the six nearest neighbours is more than 20~pc away, we exclude the star from our sample to maintain spatial coherence.

The extinction values ($E$) from \citet{Zhang2025} are converted to \textit{Gaia} broadband $G$, $G_{\rm BP}$, and $G_{\rm RP}$ extinctions using the \texttt{dustapprox} package \citep{dustapprox_2022}. This package requires the extinction at 550~nm ($A_0$) and the effective temperature ($T_{\rm eff}$). We compute $A_0$ using $A_0 = R \cdot E$, with $R = 2.69$ \citep{zhang2023, Zhang2025}. Given our focus on YSOs, we adopt a uniform effective temperature of $T_{\rm eff} = 5000\,\mathrm{K}$ for all stars in the sample to obtain \textit{Gaia} broadband extinctions. To assess the sensitivity of this choice, we also perform the conversion using $T_{\rm eff} = 3000\,\mathrm{K}$ and find negligible differences: the typical variations in $E(G_{\rm BP} - G_{\rm RP})$ and $A_G$ are only 0.01 and 0.02~mag, respectively. After applying distance and extinction corrections, we construct the HR diagram for our sample, as shown in the left-hand panel of Fig.~\ref{fig:hrd}.

\subsection{Preliminary Selection of YSO Candidates}
To identify YSO candidates, we analyse the HR diagram using the PARSEC~1.2S isochrones \citep{bressan2012, chen2014, tang2014, chen2015}, adopting a solar metallicity ([Fe/H] = 0~dex) and an age of 20~Myr. We further restrict the sample to stars with absolute magnitudes $M_G > 4~\mathrm{mag}$ and $M_G < 11~\mathrm{mag}$ to exclude giant stars and very cool stars \citep{zari2018, Anderson2025}. To mitigate the influence of binary systems, we follow \citet{zari2018} and define a binary sequence, which is brighter than the main sequence (based on PARSEC isochrone with [Fe/H] = 0~dex and an age of 1 Gyr) by 0.75~mag. We then select stars that are brighter than this binary sequence. By applying these criteria, we obtain a refined sample of $\sim$ 740{,}000 YSO candidates, as illustrated in the right-hand panel of Fig.~\ref{fig:hrd}. We note that this selection based on the colour-magnitude diagram (CMD) is necessarily imprecise. Photometric errors, distance uncertainties, and extinction corrections introduce scatter into the HR diagram, which broadens the observed main sequence and PMS region. As a result, the initial candidate sample is subject to contamination from field stars, in particular unresolved binary systems that scatter above the main sequence. Field stars vastly outnumber genuine YSOs in the Milky Way, and the right panel of Fig.~\ref{fig:hrd} shows that the stellar density increases markedly towards the main sequence, where a significant fraction of the sources are likely contaminants rather than true young stars. This contamination motivates the subsequent HDBSCAN clustering step, which filters the sample by requiring statistically significant density enhancements in the six-dimensional phase space.

\begin{figure*}
	\includegraphics[width=0.58\textwidth]{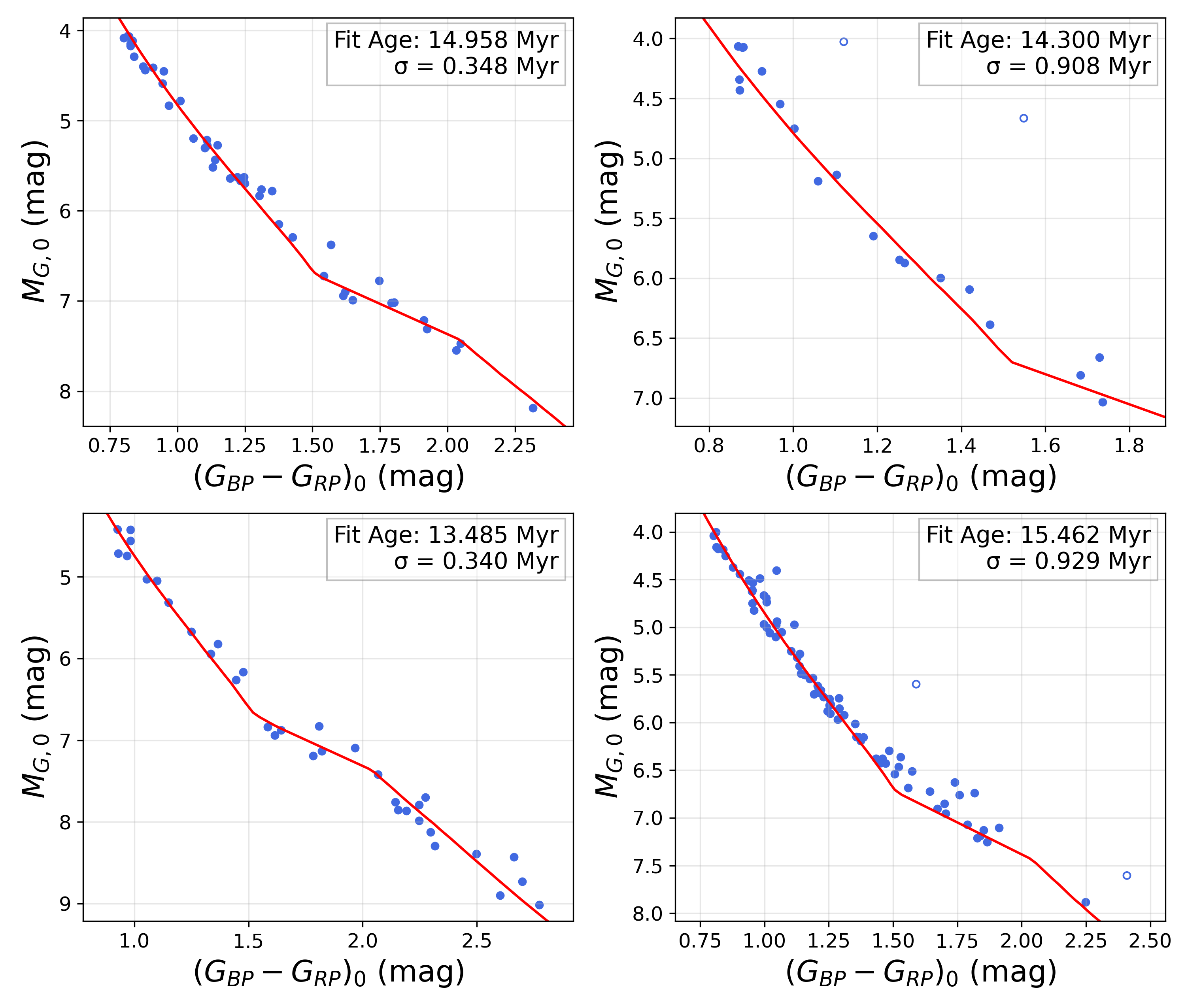}
    \caption{Illustration of isochrone fitting for YSO association ages. Blue points represent member stars of a stellar group, and the red solid line indicates the best-fitting isochrone model. Open circles indicate outliers that are not included in the age fitting.}
    \label{fig:age_fit}
\end{figure*}

\begin{figure*}
	\includegraphics[width=\textwidth]{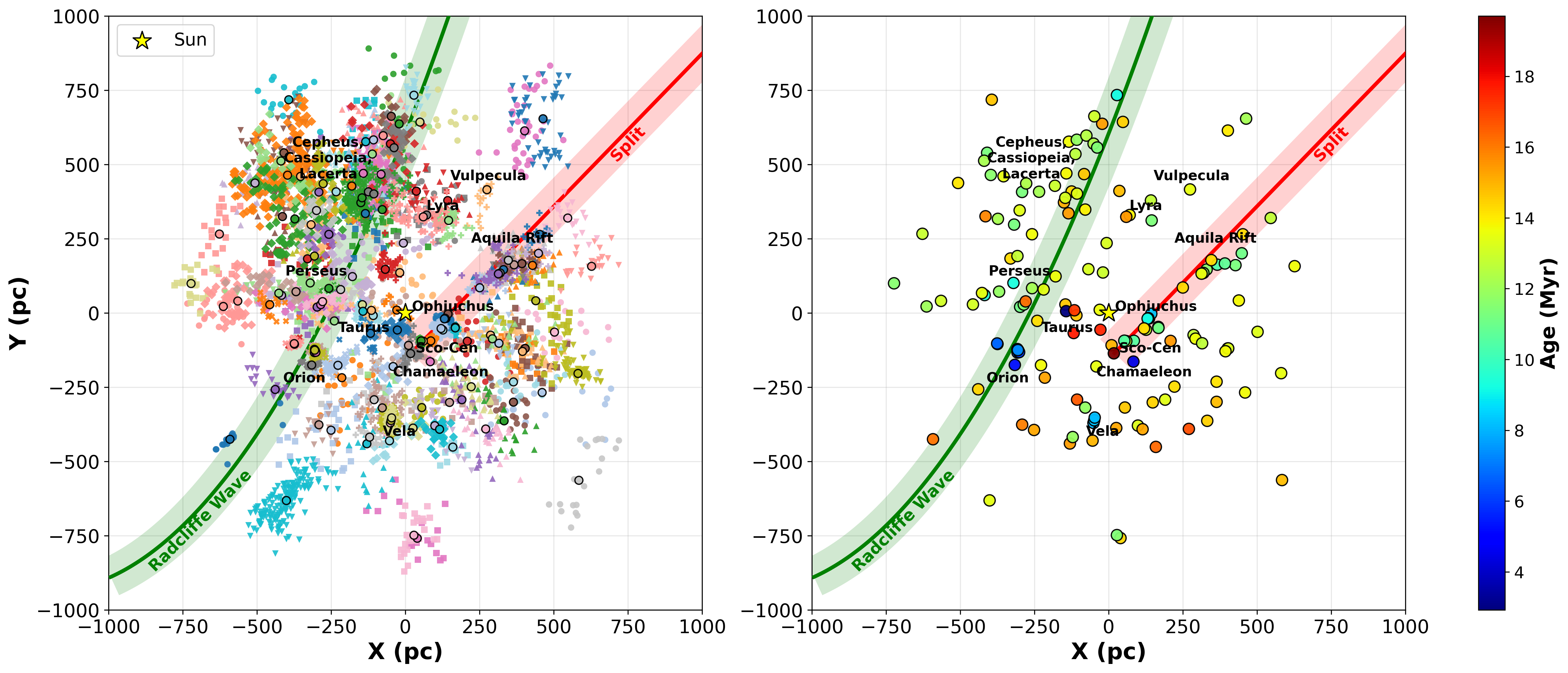}
    \caption{Spatial distribution of our identified YSO associations in the Galactic $X$--$Y$ plane. Left panel: member stars shown with different colours and marker styles according to their HDBSCAN group membership. Filled circles with black edges mark the centre of each group, shown in the same colour as its member stars. Right panel: group centres only, colour-coded by isochrone-fitted age (colour bar in Myr). In both panels, the yellow star marks the position of the Sun. Black labels indicate major molecular clouds and star-forming regions in the solar neighbourhood. The green curve and shaded region show the projected location and approximate spatial extent of the Radcliffe Wave, and the red line and shaded region show the Split.}
    \label{fig:yso_xy}
\end{figure*}

\subsection{Clustering and Parameter Estimation of YSO Associations}
For clustering YSO associations, we use the Cartesian spatial coordinates $(x, y, z)$ and the three velocity components $(V_{\alpha}, V_{\delta}, V_r)$ of each star as input parameters. Here, $x$ is directed from the Sun towards the Galactic centre, $y$ follows the direction of Galactic rotation, and $z$ is perpendicular to the Galactic plane towards the north Galactic pole. The velocity components $V_{\alpha}$ and $V_{\delta}$ represent tangential velocities along right ascension and declination, respectively, while $V_r$ is the radial velocity along the line of sight. To compute the tangential velocities, we propagate measurement uncertainties using a Monte Carlo approach. For proper motions $(\mu_{\alpha}, \mu_{\delta})$ and distances ($d$), we perform 1000 Gaussian samplings based on their uncertainties, calculate $V_{\alpha}$ and $V_{\delta}$ for each sampling, and adopt the mean of these values as the final velocities, with the standard deviations as their uncertainties.

For clustering, we employ the HDBSCAN algorithm \citep{mcinnes2017}, which is widely used in previous studies for identifying stellar groups in the Milky Way due to its effectiveness in handling datasets with varying densities \citep{kounkel2019, kounkel2020a, hunt2023, hunt2024}. HDBSCAN evaluates all possible DBSCAN clustering solutions across the dataset and selects optimal clusters based on a minimum cluster size, defined by the parameter \texttt{min\_cluster\_size}. Prior to standardisation, we restrict the total velocity of stars to $v_{\rm total} = \sqrt{V_{\alpha}^2 + V_{\delta}^2 + V_r^2} < 60~\mathrm{km\,s^{-1}}$ \citep{bensby2014, zari2018} to select disc stars, exclude halo stars, and remove velocity outliers. Since we perform clustering in a six-dimensional phase space, we standardise the data using the $z$-score method (which uses $z = \frac{x - \mu}{\sigma}$ to transform the data to a standard distribution with a mean of 0 and a standard deviation of 1) to account for differences in units between spatial coordinates $(x, y, z)$ and velocity components $(V_{\alpha}, V_{\delta}, V_r)$. The standard deviations of the spatial coordinates are $\sigma_x = 0.44~\mathrm{kpc}$, $\sigma_y = 0.44~\mathrm{kpc}$, and $\sigma_z = 0.26~\mathrm{kpc}$, while those of the velocity components are $\sigma_{V_\alpha} = 20.85~\mathrm{km\,s^{-1}}$, $\sigma_{V_\delta} = 20.17~\mathrm{km\,s^{-1}}$, and $\sigma_{V_r} = 24.03~\mathrm{km\,s^{-1}}$. The standardised data are used only for clustering, while all subsequent analyses are based on the original, unstandardised data. We note that the precision of Gaia DR3 radial velocities is generally lower than that of the tangential velocities derived from proper motions. However, because all six phase-space coordinates are standardised via $z$-score normalisation prior to clustering, the radial velocity dimension is not disproportionately weighted relative to the spatial or tangential velocity dimensions. More importantly, measurement uncertainties act to broaden the distribution of stars in phase space rather than to artificially concentrate them. Larger radial velocity uncertainties primarily reduce the density contrast of genuine associations along the line-of-sight direction, thereby lowering the recovery efficiency for peripheral members or low-density, diffuse associations. They are unlikely to produce spurious compact clusters, as HDBSCAN identifies structures based on statistically significant local density enhancements, which random measurement scatter does not generate. Furthermore, the subsequent isochrone fitting and visual inspection of the HR diagram provide additional astrophysical consistency checks that help filter out any kinematic false positives \citep[see also][]{hunt2023}.

After testing various HDBSCAN parameters, we adopt \texttt{min\_samples} = 5, \texttt{min\_cluster\_size} = 20, and the \texttt{leaf} clustering mode to achieve optimal results. This process yields 161 YSO groups comprising 6916 member stars. The spatial distribution of these member stars in Galactic coordinates is shown in Fig.~\ref{fig:yso_lb}.

After identifying the YSO groups, we estimate their ages by fitting isochrones. We use the PARSEC isochrone models \citep{bressan2012, tang2014, chen2014, chen2015}, adopting solar metallicity ([Fe/H] = 0~dex) and generating 200 isochrone models spanning ages from 0.1 to 20~Myr at 0.1~Myr intervals. For each group, we calculate the chi-squared distance between the member stars’ positions on the HR diagram and each isochrone model for each group, employing a distance weighting and excluding outliers that lie beyond 0.25~mag from the model. We select the 10 models with the smallest chi-squared values, fit the relationship between chi-squared and age, and interpolate to determine the age corresponding to the minimum chi-squared value. This age is adopted as the group’s age. To estimate age uncertainties, we use a Monte Carlo approach, generating 1000 datasets for each group by sampling the absolute magnitudes and intrinsic colours of member stars based on their uncertainties. Each dataset is fitted with isochrones to obtain 1000 age measurements, with the mean adopted as the group’s age and the standard deviation as the age uncertainty $\sigma_{\rm age}$. Examples of age fits for selected YSO associations are shown in Fig.~\ref{fig:age_fit}.

In addition to age, we calculate the velocity dispersion $\sigma_v$ and physical size $r$ for each YSO group, following the methodology of \citet{zhou2022}. For velocity dispersion, we use a Monte Carlo approach, generating 1000 datasets for each group by sampling the velocity components based on their uncertainties. Each dataset yields a velocity dispersion value, and the mean of the 1000 values is adopted as $\sigma_v$, with the standard deviation as the uncertainty $\sigma_{v,\mathrm{error}}$. For the group size, we fit an ellipsoidal model to the distribution of member stars in $(x, y, z)$ coordinates. The radius $r$ is expressed in parsecs. To account for distance uncertainties, we generate 1000 datasets for each group by sampling the spatial coordinates based on distance errors. The mean of the 1000 radius values is adopted as the group’s radius $r$, with the standard deviation as the uncertainty $r_{\mathrm{error}}$.

\begin{figure}
	\centering
	\includegraphics[width=\columnwidth]{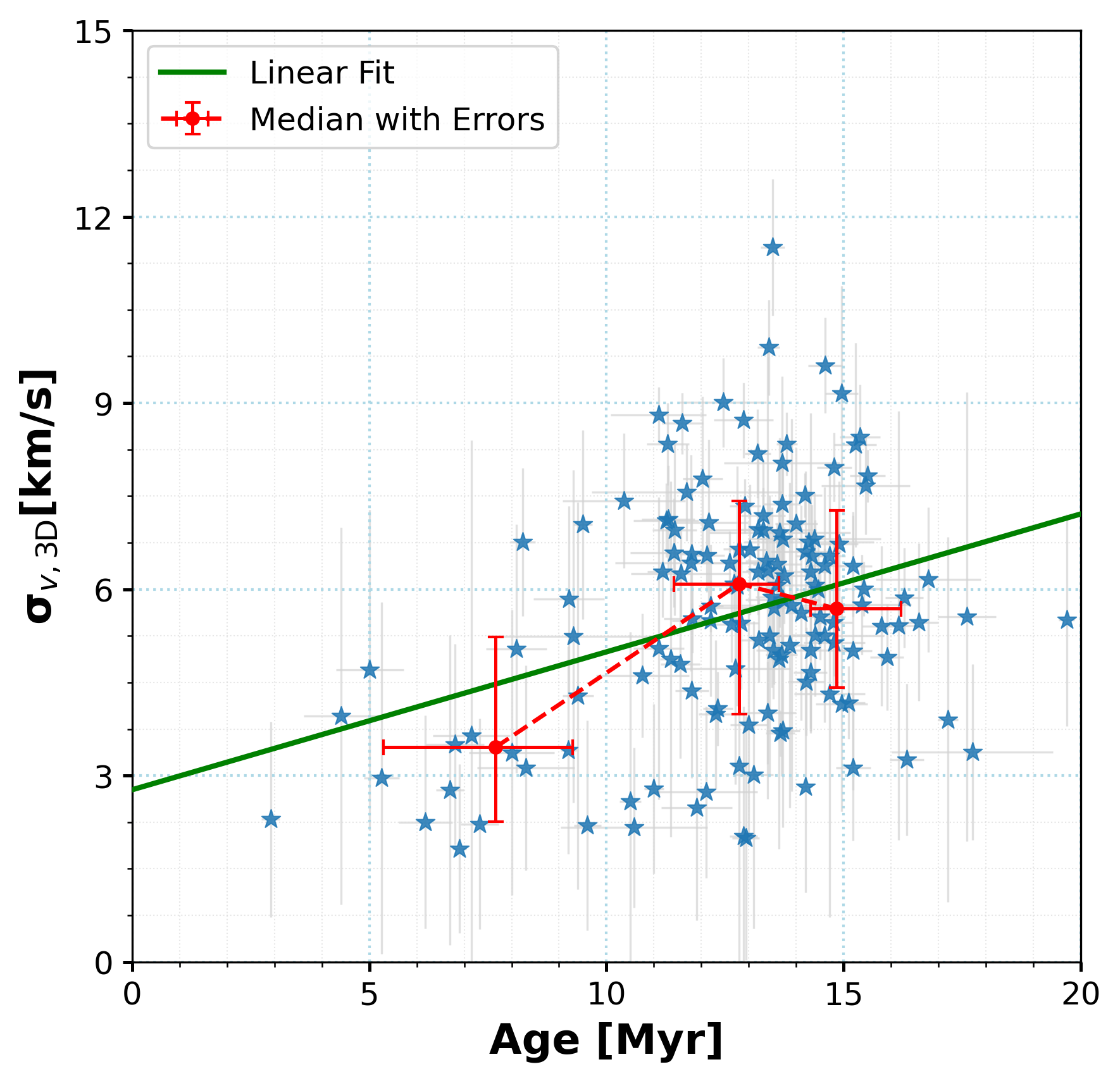}
    \caption{Relationship between velocity dispersion and age for YSO associations.
	Blue stars represent individual groups, with grey error bars indicating age and
	velocity dispersion uncertainties. Red circles mark the median velocity dispersion
	for different age bins, connected by a red dashed line, while the green solid line
	shows the linear fit to the data.}
    \label{fig:sigma_v_age}
\end{figure}

\begin{figure*}
	\centering
	\includegraphics[width=\textwidth]{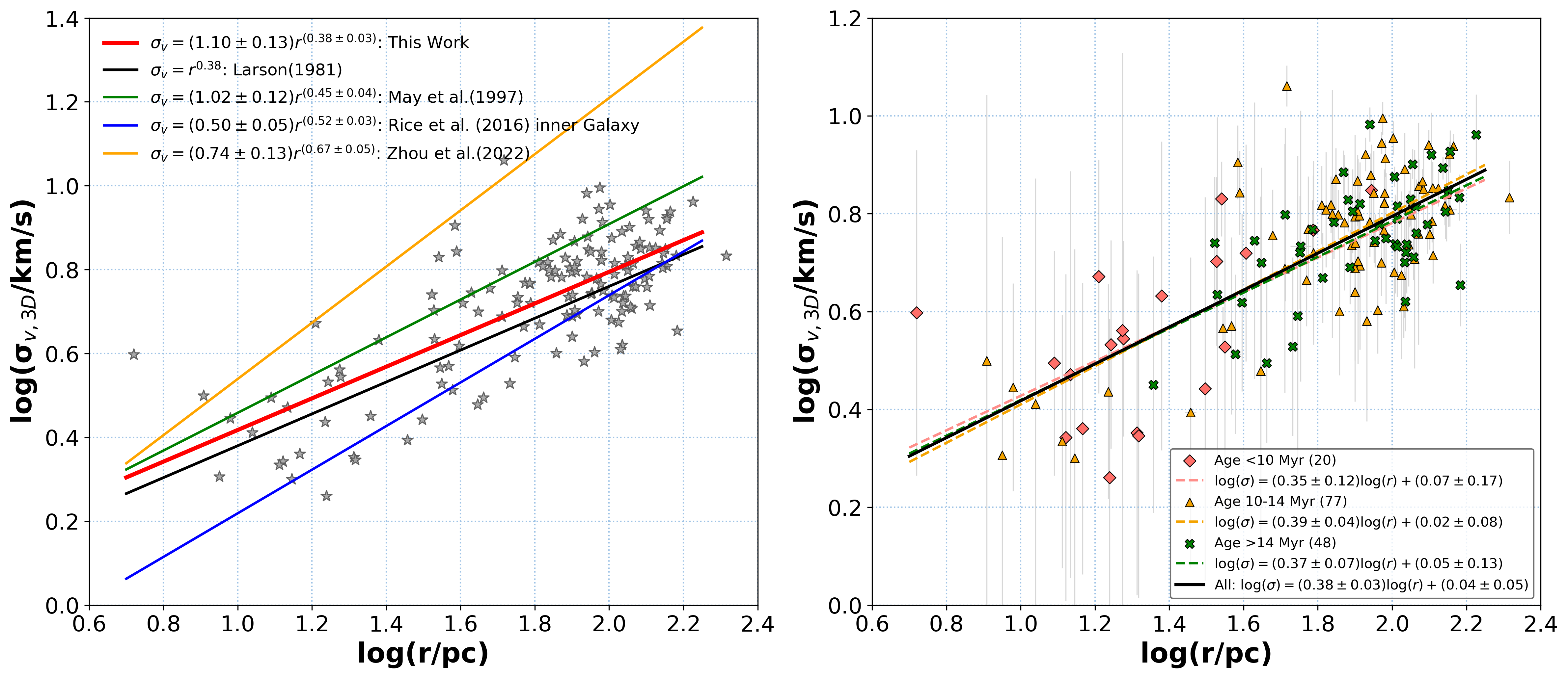}
    \caption{Left panel: Relationship between velocity dispersion and radius for YSO associations. Grey stars represent our sample groups. The orange solid line shows the Larson relation derived from young stars' two-dimensional velocities \citep{zhou2022}, the green solid line represents the relation from southern molecular clouds' CO data \citep{may1997}, the black solid line corresponds to \citet{larson1981}, the red solid line is our fitted relation, and the blue solid line represents the relation from inner Galactic molecular clouds' CO data \citep{thomas2016}. Right panel: Larson's relation for different age groups. Red squares represent groups younger than 10~Myr, orange triangles denote groups aged 10--14~Myr, and green crosses indicate groups older than 14~Myr. The black solid line is the fit for the entire sample, while the red, orange, and green dashed lines represent fits for the respective age groups. Grey thin lines indicate data uncertainties.}
    \label{fig:sigma_v_r}
\end{figure*}

\section{Results and Discussion}
Using the HR diagram derived from \textit{Gaia} data, we identify YSO candidates and perform HDBSCAN clustering in six-dimensional phase space, resulting in 161 stellar groups with 6916 member stars. During the age estimation process, we first fit a best-matching isochrone to each HDBSCAN group on the extinction-corrected HR diagram. Individual member stars whose photometric distance from the best-fitting isochrone exceeds 0.25~mag are flagged as outliers and excluded from the age determination. We then visually inspect the HR diagram of each group together with its best-fitting isochrone. Groups whose member stars do not collectively form a clear sequence consistent with a single isochrone are removed from the final age-fitting sample. This results in a final sample of 145 groups comprising 5713 member stars\footnote{The catalogues of our 145 identified YSO groups are publicly accessible at \url{https://nadc.china-vo.org/res/r101835/}.}. In the following analysis, we focus only on the sample with robust age measurements. To further assess the reliability of our YSO association catalogue, we have cross-matched our sample with published young-star catalogues; the results are presented in Appendix~\ref{sec:comparison_with_published_young_star_samples}.

The spatial distribution of these YSO associations in the Galactic plane is shown in Fig.~\ref{fig:yso_xy}. Most associations are located within 1~kpc of the Sun, primarily due to the brightness of \textit{Gaia} stars with RV measurements. The spatial distribution of our YSO associations aligns well with previous studies \citep{perrot2003, pm2016, zari2018, zucker2022, Zucker2023}, reproducing known structures in the solar neighbourhood and confirming the reliability of our results. Notable star-forming regions, such as Orion, Perseus, Scorpius--Centaurus, and Vela, are evident in the distribution. The left panel of Fig.~\ref{fig:yso_xy} shows that the distribution of our YSO associations exhibits a degree of correspondence with large-scale structures in the solar neighbourhood, including the Radcliffe Wave \citep{alves2020, li2022} and the Split \citep{lallement2019, Chen2020}. The right panel, in which group centres are colour-coded by their isochrone-fitted ages, further indicates that the youngest associations tend to be located near the loci of these structures. While overall no strong global age--position trend is apparent, the youngest groups ($<$\,10~Myr) show a noticeable concentration towards the Radcliffe Wave and Split, consistent with the picture that these large-scale gas structures serve as reservoirs for recent star formation \citep{alves2020, Zucker2023, Kormann2026}. Beyond these structures, the spatial distribution of our YSO associations may also be related to the local superclouds recently identified by \citet{Kormann2026} from Gaia-based 3D dust maps. A qualitative comparison indicates that our YSO associations show broad spatial consistency with several nearby superclouds, including the Radcliffe Wave Cloud, Vela Ridge Cloud, Malpolon Cloud, Natrix Cloud, and the Split.

\subsection{Relationship Between Velocity Dispersion and Age}
Our sample of YSO associations, with measurements of age, three-dimensional positions, and velocities, provides an ideal dataset for studying the kinematic evolution of these groups. We first examine the relationship between velocity dispersion and age, as illustrated in Fig.~\ref{fig:sigma_v_age}. 

To explore this relationship, we divide the groups into three age bins: younger than 10~Myr, between 10 and 14~Myr inclusive, and older than 14~Myr. As shown in Fig.~\ref{fig:sigma_v_age}, both the median velocity dispersion across age bins and the linear fit of all the 145 YSO associations indicate a positive correlation, with velocity dispersion increasing as groups age. This trend is expected, as velocity dispersion typically grows due to internal dynamical evolution and external Galactic heating \citep{larson1981, solomon1987}. Our results confirm this pattern, showing that older groups exhibit larger velocity dispersions.

\subsection{Relationship Between Velocity Dispersion and Radius}
\citet{larson1981} proposed that the velocity dispersion ($\sigma_v$) of molecular clouds scales with their size ($r$) as $\sigma_v \propto r^{\beta}$, with $\beta = 0.38$, a relationship known as Larson's law. While the form is widely accepted, the value of $\beta$ remains debated. Larson's relation is expressed as:
\begin{equation}
\sigma_v = A \times r^{\beta},
\end{equation}
where $A$ is a proportionality constant and $\beta$ is the power-law index. 

We plot the velocity dispersion versus radius for our YSO associations in the left-hand panel of Fig.~\ref{fig:sigma_v_r} and derive the relation $\sigma_v = (1.10 \pm 0.13) \times r^{0.38 \pm 0.03}$. The fitted slope is consistent with the typical range of 0.4--0.5 and aligns with the value of 0.38 proposed by \citet{larson1981}.

Previous studies have reported varying $\beta$ values. For instance, \citet{may1997} analysed CO observations of 177 southern molecular clouds, finding $\sigma_v = (1.02 \pm 0.12) \times r^{0.45 \pm 0.04}$, with $\beta = 0.45 \pm 0.04$. Similarly, \citet{thomas2016} studied 1064 molecular clouds across the sky, deriving $\sigma_v = (0.5 \pm 0.05) \times r^{0.52 \pm 0.03}$ for Galactic clouds, with $\beta = 0.52 \pm 0.03$. Conversely, \citet{caselli1995} found $\beta = 0.53 \pm 0.07$ for low-mass cloud cores in Orion~A, consistent with typical values, but $\beta = 0.21 \pm 0.03$ for high-mass cores, deviating significantly from 0.38. \citet{huang2023} reported $\beta = 0.20 \pm 0.12$ for high-mass star-forming regions, and $\beta = 0.18 \pm 0.15$ when excluding Galactic centre samples, attributing deviations to complex dynamics and magnetic fields inhibiting collapse in small, dense regions. Higher $\beta$ values have also been reported, such as $\beta = 0.67 \pm 0.05$ by \citet{zhou2022} using two-dimensional velocities of young stars and $\beta = 0.66 \pm 0.003$ by \citet{sun2024} for molecular clouds in the outer Galactic disc. 

These variations reflect the complex turbulent dynamics of molecular clouds across different environments, influenced by factors such as internal magnetic fields, stellar feedback, external gravitational fields, and kinetic energy injection. Our derived $\beta = 0.38$ supports the traditional view that turbulence dominates during molecular cloud collapse, though other mechanisms likely contribute to the overall evolution.

\subsection{Larson’s Relation Across Age Groups}
Given that our YSO associations have age measurements, we investigate whether Larson's relation evolves with time by calculating the relation for three age-based subsamples: groups younger than 10~Myr, 10--14~Myr, and older than 14~Myr. The velocity dispersion versus radius relationships for these subsamples are shown in the right-hand panel of Fig.~\ref{fig:sigma_v_r}. 

The fitted Larson relations for the age groups yield $\beta$ values of $0.35 \pm 0.12$, $0.39 \pm 0.04$, and $0.37 \pm 0.07$, respectively, all conforming to the typical range of 0.4--0.5. The near-parallel fits suggest that Larson's relation does not significantly evolve with time, indicating consistent turbulent dynamics across different age groups in our sample. 

The lack of evolution in Larson's relation across different age groups of young stellar object (YSO) associations suggests that the kinematic properties, inherited from the turbulent dynamics of their parent molecular clouds, remain stable over the age range up to 20~Myr. This consistency, with $\beta$ values of 0.35--0.39 across age bins, indicates that the scaling relation between velocity dispersion and size, as described by $\sigma_v \propto r^{\beta}$ \citep{larson1981}, is largely preserved without significant disruption from internal dynamical processes. The stability suggests that the initial turbulent structure of the parent clouds dominates the kinematic characteristics of these groups, with minimal alteration from secondary processes such as stellar feedback or magnetic fields over the studied timescale. This finding reinforces the idea that the kinematic properties of YSO associations are primarily a legacy of their formation environment, providing a benchmark for understanding the persistence of turbulent signatures in young stellar systems.

\section{Conclusions}

This study leverages high-precision photometry, astrometry, and radial velocity data from \textit{Gaia} DR3 to identify YSO candidates through the HR diagram. By combining six-dimensional data $(x, y, z, V_{\alpha}, V_{\delta}, V_r)$ and applying HDBSCAN clustering, we identify 145 YSO groups comprising 5713 stars within 1~kpc of the Sun. We measure key parameters for these groups, including age, velocity dispersion, and radius, providing a robust sample for studying the kinematic properties and evolution of young stellar groups. Using this sample, we investigate the relationship between velocity dispersion and radius, deriving the relation $\sigma_v = (1.10 \pm 0.13) \times r^{0.38 \pm 0.03}$. The fitted slope is consistent with the typical range of 0.4--0.5 and aligns with the value of 0.38 proposed by \citet{larson1981}. For the first time, we examine Larson's relation across different age groups, finding similar slopes for all age bins. This consistency indicates that Larson's relation does not evolve significantly with time, demonstrating its stability over the age range studied. 

The findings of this study advance our understanding of the kinematic and evolutionary properties of young stellar groups, providing a robust dataset for probing the dynamics of star-forming regions. By confirming the stability of Larson's relation across different age groups, our work highlights that the turbulent properties inherited from parent molecular clouds persist without significant disruption, offering a benchmark for studying the kinematic legacy of stellar group formation. Looking ahead, we plan to extend this study beyond the current Gaia DR3 RV-selected YSO sample by incorporating data from spectroscopic surveys such as LAMOST and APOGEE. This expansion will allow us to analyze older stellar tracers and more distant objects, to test whether the kinematic scaling relations found here for YSO associations persist in more evolved and distant young stellar populations, and to investigate how inherited turbulent structures interact with secondary processes such as magnetic fields and stellar feedback during the evolution of young stellar associations.

\section*{Acknowledgements}

We thank the anonymous referee for a thorough and constructive report that helped us improve the clarity and rigour of this work.
This work is partially supported by the National Natural Science Foundation of China (12173034 and 12322304). We acknowledge the science research grants from the China Manned Space Project with No.~CMS-CSST-2025-A11. 

This work presents results from the European Space Agency (ESA) space mission \textit{Gaia}. \textit{Gaia} data are being processed by the \textit{Gaia} Data Processing and Analysis Consortium (DPAC). Funding for the DPAC is provided by national institutions, in particular the institutions participating in the \textit{Gaia} MultiLateral Agreement (MLA). The \textit{Gaia} mission website is \url{https://www.cosmos.esa.int/gaia}. The \textit{Gaia} archive website is \url{https://archives.esac.esa.int/gaia}.

\section*{Data Availability}
The catalogues of our 145 identified YSO groups can be accessed at \url{https://nadc.china-vo.org/res/r101835/}.



\bibliographystyle{mnras}
\bibliography{gaiayso} 

@ARTICLE{Chen2019OBStar,
       author = {{Chen}, B. -Q. and {Huang}, Y. and {Hou}, L. -G. and {Tian}, H. and {Li}, G. -X. and {Yuan}, H. -B. and {Wang}, H. -F. and {Wang}, C. and {Tian}, Z. -J. and {Liu}, X. -W.},
        title = "{The Galactic spiral structure as revealed by O- and early B-type stars}",
      journal = {\mnras},
         year = 2019,
        month = jul,
       volume = {487},
       number = {1},
        pages = {1400-1409},
          doi = {10.1093/mnras/stz1357},
archivePrefix = {arXiv},
       eprint = {1905.05542},
 primaryClass = {astro-ph.GA},
       adsurl = {https://ui.adsabs.harvard.edu/abs/2019MNRAS.487.1400C}
}

@ARTICLE{BailerJones2018,
       author = {{Bailer-Jones}, C.~A.~L. and {Rybizki}, J. and {Fouesneau}, M. and {Mantelet}, G. and {Andrae}, R.},
        title = "{Estimating Distance from Parallaxes. IV. Distances to 1.33 Billion Stars in Gaia Data Release 2}",
      journal = {AJ},
         year = 2018,
        month = aug,
       volume = {156},
       number = {2},
          eid = {58},
        pages = {58},
          doi = {10.3847/1538-3881/aacb21},
archivePrefix = {arXiv},
       eprint = {1804.10121},
 primaryClass = {astro-ph.SR},
       adsurl = {https://ui.adsabs.harvard.edu/abs/2018AJ....156...58B}
}

@ARTICLE{Shen2022,
       author = {{Shen}, H. and {Chen}, B. -Q. and {Guo}, H. -L. and {Yuan}, H. -B. and {Sun}, W. -X. and {Li}, J.},
        title = "{Estimating accurate reddening values of LAMOST M dwarfs}",
      journal = {\mnras},
         year = 2022,
        month = aug,
       volume = {514},
       number = {3},
        pages = {4398-4405},
          doi = {10.1093/mnras/stac1615},
archivePrefix = {arXiv},
       eprint = {2206.03632},
 primaryClass = {astro-ph.SR},
       adsurl = {https://ui.adsabs.harvard.edu/abs/2022MNRAS.514.4398S}
}

@ARTICLE{Chen2017,
       author = {{Chen}, B. -Q. and {Liu}, X. -W. and {Yuan}, H. -B. and {Robin}, A.~C. and {Huang}, Y. and {Xiang}, M. -S. and {Wang}, C. and {Ren}, J. -J. and {Tian}, Z. -J. and {Zhang}, H. -W.},
        title = "{Constraining the Galactic structure parameters with the XSTPS-GAC and SDSS photometric surveys}",
      journal = {\mnras},
         year = 2017,
        month = jan,
       volume = {464},
       number = {3},
        pages = {2545-2556},
          doi = {10.1093/mnras/stw2497},
archivePrefix = {arXiv},
       eprint = {1609.08838},
 primaryClass = {astro-ph.GA},
       adsurl = {https://ui.adsabs.harvard.edu/abs/2017MNRAS.464.2545C}
}

@ARTICLE{zari2018,
       author = {{Zari}, E. and {Hashemi}, H. and {Brown}, A.~G.~A. and {Jardine}, K. and {de Zeeuw}, P.~T.},
        title = "{3D mapping of young stars in the solar neighbourhood with Gaia DR2}",
      journal = {A\&A},
         year = 2018,
        month = dec,
       volume = {620},
          eid = {A172},
        pages = {A172},
          doi = {10.1051/0004-6361/201834150},
archivePrefix = {arXiv},
       eprint = {1810.09819},
 primaryClass = {astro-ph.SR},
       adsurl = {https://ui.adsabs.harvard.edu/abs/2018A&A...620A.172Z}
}

@article{zucker2022,
	adsurl = {https://ui.adsabs.harvard.edu/abs/2022Natur.601..334Z},
	archiveprefix = {arXiv},
	author = {{Zucker}, Catherine and {Goodman}, Alyssa A. and {Alves}, Jo{\~a}o and {Bialy}, Shmuel and {Foley}, Michael and {Speagle}, Joshua S. and {Gro{\^I}{\texttwosuperior}schedl}, Josefa and {Finkbeiner}, Douglas P. and {Burkert}, Andreas and {Khimey}, Diana and {Swiggum}, Cameren},
	doi = {10.1038/s41586-021-04286-5},
	eprint = {2201.05124},
	journal = {Nature},
	month = jan,
	number = {7893},
	pages = {334-337},
	primaryclass = {astro-ph.GA},
	title = {{Star formation near the Sun is driven by expansion of the Local Bubble}},
	volume = {601},
	year = 2022}

@article{kounkel2019,
	adsurl = {https://ui.adsabs.harvard.edu/abs/2019AJ....158..122K},
	archiveprefix = {arXiv},
	author = {{Kounkel}, Marina and {Covey}, Kevin},
	doi = {10.3847/1538-3881/ab339a},
	eid = {122},
	eprint = {1907.07709},
	journal = {AJ},
	month = sep,
	number = {3},
	pages = {122},
	primaryclass = {astro-ph.GA},
	title = {{Untangling the Galaxy. I. Local Structure and Star Formation History of the Milky Way}},
	volume = {158},
	year = 2019}

@article{gaia2023,
	adsurl = {https://ui.adsabs.harvard.edu/abs/2023A&A...674A...1G},
	archiveprefix = {arXiv},
	author = {{Gaia Collaboration} and {Vallenari}, A. and {Brown}, A.~G.~A. and {Prusti}, T. and {de Bruijne}, J.~H.~J. and {Arenou}, F. and {Babusiaux}, C. and {Biermann}, M. and {Creevey}, O.~L. and {Ducourant}, C. and {Evans}, D.~W. and {Eyer}, L. and {Guerra}, R. and {Hutton}, A. and {Jordi}, C. and {Klioner}, S.~A. and {Lammers}, U.~L. and {Lindegren}, L. and {Luri}, X. and {Mignard}, F. and {Panem}, C. and {Pourbaix}, D. and {Randich}, S. and {Sartoretti}, P. and {Soubiran}, C. and {Tanga}, P. and {Walton}, N.~A. and {Bailer-Jones}, C.~A.~L. and {Bastian}, U. and {Drimmel}, R. and {Jansen}, F. and {Katz}, D. and {Lattanzi}, M.~G. and {van Leeuwen}, F. and {Bakker}, J. and {Cacciari}, C. and {Casta{\~n}eda}, J. and {De Angeli}, F. and {Fabricius}, C. and {Fouesneau}, M. and {Fr{\'e}mat}, Y. and {Galluccio}, L. and {Guerrier}, A. and {Heiter}, U. and {Masana}, E. and {Messineo}, R. and {Mowlavi}, N. and {Nicolas}, C. and {Nienartowicz}, K. and {Pailler}, F. and {Panuzzo}, P. and {Riclet}, F. and {Roux}, W. and {Seabroke}, G.~M. and {Sordo}, R. and {Th{\'e}venin}, F. and {Gracia-Abril}, G. and {Portell}, J. and {Teyssier}, D. and {Altmann}, M. and {Andrae}, R. and {Audard}, M. and {Bellas-Velidis}, I. and {Benson}, K. and {Berthier}, J. and {Blomme}, R. and {Burgess}, P.~W. and {Busonero}, D. and {Busso}, G. and {C{\'a}novas}, H. and {Carry}, B. and {Cellino}, A. and {Cheek}, N. and {Clementini}, G. and {Damerdji}, Y. and {Davidson}, M. and {de Teodoro}, P. and {Nu{\~n}ez Campos}, M. and {Delchambre}, L. and {Dell'Oro}, A. and {Esquej}, P. and {Fern{\'a}ndez-Hern{\'a}ndez}, J. and {Fraile}, E. and {Garabato}, D. and {Garc{\'\i}a-Lario}, P. and {Gosset}, E. and {Haigron}, R. and {Halbwachs}, J. -L. and {Hambly}, N.~C. and {Harrison}, D.~L. and {Hern{\'a}ndez}, J. and {Hestroffer}, D. and {Hodgkin}, S.~T. and {Holl}, B. and {Jan{\ss}en}, K. and {Jevardat de Fombelle}, G. and {Jordan}, S. and {Krone-Martins}, A. and {Lanzafame}, A.~C. and {L{\"o}ffler}, W. and {Marchal}, O. and {Marrese}, P.~M. and {Moitinho}, A. and {Muinonen}, K. and {Osborne}, P. and {Pancino}, E. and {Pauwels}, T. and {Recio-Blanco}, A. and {Reyl{\'e}}, C. and {Riello}, M. and {Rimoldini}, L. and {Roegiers}, T. and {Rybizki}, J. and {Sarro}, L.~M. and {Siopis}, C. and {Smith}, M. and {Sozzetti}, A. and {Utrilla}, E. and {van Leeuwen}, M. and {Abbas}, U. and {{\'A}brah{\'a}m}, P. and {Abreu Aramburu}, A. and {Aerts}, C. and {Aguado}, J.~J. and {Ajaj}, M. and {Aldea-Montero}, F. and {Altavilla}, G. and {{\'A}lvarez}, M.~A. and {Alves}, J. and {Anders}, F. and {Anderson}, R.~I. and {Anglada Varela}, E. and {Antoja}, T. and {Baines}, D. and {Baker}, S.~G. and {Balaguer-N{\'u}{\~n}ez}, L. and {Balbinot}, E. and {Balog}, Z. and {Barache}, C. and {Barbato}, D. and {Barros}, M. and {Barstow}, M.~A. and {Bartolom{\'e}}, S. and {Bassilana}, J. -L. and {Bauchet}, N. and {Becciani}, U. and {Bellazzini}, M. and {Berihuete}, A. and {Bernet}, M. and {Bertone}, S. and {Bianchi}, L. and {Binnenfeld}, A. and {Blanco-Cuaresma}, S. and {Blazere}, A. and {Boch}, T. and {Bombrun}, A. and {Bossini}, D. and {Bouquillon}, S. and {Bragaglia}, A. and {Bramante}, L. and {Breedt}, E. and {Bressan}, A. and {Brouillet}, N. and {Brugaletta}, E. and {Bucciarelli}, B. and {Burlacu}, A. and {Butkevich}, A.~G. and {Buzzi}, R. and {Caffau}, E. and {Cancelliere}, R. and {Cantat-Gaudin}, T. and {Carballo}, R. and {Carlucci}, T. and {Carnerero}, M.~I. and {Carrasco}, J.~M. and {Casamiquela}, L. and {Castellani}, M. and {Castro-Ginard}, A. and {Chaoul}, L. and {Charlot}, P. and {Chemin}, L. and {Chiaramida}, V. and {Chiavassa}, A. and {Chornay}, N. and {Comoretto}, G. and {Contursi}, G. and {Cooper}, W.~J. and {Cornez}, T. and {Cowell}, S. and {Crifo}, F. and {Cropper}, M. and {Crosta}, M. and {Crowley}, C. and {Dafonte}, C. and {Dapergolas}, A. and {David}, M. and {David}, P. and {de Laverny}, P. and {De Luise}, F. and {De March}, R.},
	doi = {10.1051/0004-6361/202243940},
	eid = {A1},
	eprint = {2208.00211},
	journal = {A\&A},
	month = jun,
	pages = {A1},
	primaryclass = {astro-ph.GA},
	title = {{Gaia Data Release 3. Summary of the content and survey properties}},
	volume = {674},
	year = 2023}

@article{lindegren2018,
	adsurl = {https://ui.adsabs.harvard.edu/abs/2018A&A...616A...2L},
	archiveprefix = {arXiv},
	author = {{Lindegren}, L. and {Hern{\'a}ndez}, J. and {Bombrun}, A. and {Klioner}, S. and {Bastian}, U. and {Ramos-Lerate}, M. and {de Torres}, A. and {Steidelm{\"u}ller}, H. and {Stephenson}, C. and {Hobbs}, D. and {Lammers}, U. and {Biermann}, M. and {Geyer}, R. and {Hilger}, T. and {Michalik}, D. and {Stampa}, U. and {McMillan}, P.~J. and {Casta{\~n}eda}, J. and {Clotet}, M. and {Comoretto}, G. and {Davidson}, M. and {Fabricius}, C. and {Gracia}, G. and {Hambly}, N.~C. and {Hutton}, A. and {Mora}, A. and {Portell}, J. and {van Leeuwen}, F. and {Abbas}, U. and {Abreu}, A. and {Altmann}, M. and {Andrei}, A. and {Anglada}, E. and {Balaguer-N{\'u}{\~n}ez}, L. and {Barache}, C. and {Becciani}, U. and {Bertone}, S. and {Bianchi}, L. and {Bouquillon}, S. and {Bourda}, G. and {Br{\"u}semeister}, T. and {Bucciarelli}, B. and {Busonero}, D. and {Buzzi}, R. and {Cancelliere}, R. and {Carlucci}, T. and {Charlot}, P. and {Cheek}, N. and {Crosta}, M. and {Crowley}, C. and {de Bruijne}, J. and {de Felice}, F. and {Drimmel}, R. and {Esquej}, P. and {Fienga}, A. and {Fraile}, E. and {Gai}, M. and {Garralda}, N. and {Gonz{\'a}lez-Vidal}, J.~J. and {Guerra}, R. and {Hauser}, M. and {Hofmann}, W. and {Holl}, B. and {Jordan}, S. and {Lattanzi}, M.~G. and {Lenhardt}, H. and {Liao}, S. and {Licata}, E. and {Lister}, T. and {L{\"o}ffler}, W. and {Marchant}, J. and {Martin-Fleitas}, J. -M. and {Messineo}, R. and {Mignard}, F. and {Morbidelli}, R. and {Poggio}, E. and {Riva}, A. and {Rowell}, N. and {Salguero}, E. and {Sarasso}, M. and {Sciacca}, E. and {Siddiqui}, H. and {Smart}, R.~L. and {Spagna}, A. and {Steele}, I. and {Taris}, F. and {Torra}, J. and {van Elteren}, A. and {van Reeven}, W. and {Vecchiato}, A.},
	doi = {10.1051/0004-6361/201832727},
	eid = {A2},
	eprint = {1804.09366},
	journal = {A\&A},
	month = aug,
	pages = {A2},
	primaryclass = {astro-ph.IM},
	title = {{Gaia Data Release 2. The astrometric solution}},
	volume = {616},
	year = 2018}

@article{zhang2023,
	adsurl = {https://ui.adsabs.harvard.edu/abs/2023MNRAS.524.1855Z},
	archiveprefix = {arXiv},
	author = {{Zhang}, Xiangyu and {Green}, Gregory M. and {Rix}, Hans-Walter},
	doi = {10.1093/mnras/stad1941},
	eprint = {2303.03420},
	journal = {MNRAS},
	month = sep,
	number = {2},
	pages = {1855-1884},
	primaryclass = {astro-ph.SR},
	title = {{Parameters of 220 million stars from Gaia BP/RP spectra}},
	volume = {524},
	year = 2023}

@misc{scikit-learn,
	author = {Pedregosa, F. and Varoquaux, G. and Gramfort, A. and Michel, V. and Thirion, B. and Grisel, O. and Blondel, M. and Prettenhofer, P. and Weiss, R. and Dubourg, V. and Vanderplas, J. and Passos, A. and Cournapeau, D. and Brucher, M. and Perrot, M. and Duchesnay, E.},
	note = {Accessed: 2024-10-15},
	title = {{Scikit-learn: Machine Learning in Python}},
	url = {https://scikit-learn.org/stable/modules/generated/sklearn.neighbors.NearestNeighbors.html},
	year = {2011}}

@misc{dustapprox_2022,
	author = {Fouesneau, Morgan and Andrae, Ren{\'e} and Sordo, Rosanna and Dharmawardena, Thavisha},
	month = {3},
	title = {{dustapprox}},
	url = {https://github.com/mfouesneau/dustapprox},
	version = {0.1},
	year = {2022}}

@article{bressan2012,
	adsurl = {https://ui.adsabs.harvard.edu/abs/2012MNRAS.427..127B},
	archiveprefix = {arXiv},
	author = {{Bressan}, Alessandro and {Marigo}, Paola and {Girardi}, L{\'e}o. and {Salasnich}, Bernardo and {Dal Cero}, Claudia and {Rubele}, Stefano and {Nanni}, Ambra},
	doi = {10.1111/j.1365-2966.2012.21948.x},
	eprint = {1208.4498},
	journal = {MNRAS},
	month = nov,
	number = {1},
	pages = {127-145},
	primaryclass = {astro-ph.SR},
	title = {{PARSEC: stellar tracks and isochrones with the PAdova and TRieste Stellar Evolution Code}},
	volume = {427},
	year = 2012}

@article{tang2014,
	adsurl = {https://ui.adsabs.harvard.edu/abs/2014MNRAS.445.4287T},
	archiveprefix = {arXiv},
	author = {{Tang}, Jing and {Bressan}, Alessandro and {Rosenfield}, Philip and {Slemer}, Alessandra and {Marigo}, Paola and {Girardi}, L{\'e}o and {Bianchi}, Luciana},
	doi = {10.1093/mnras/stu2029},
	eprint = {1410.1745},
	journal = {MNRAS},
	month = dec,
	number = {4},
	pages = {4287-4305},
	primaryclass = {astro-ph.SR},
	title = {{New PARSEC evolutionary tracks of massive stars at low metallicity: testing canonical stellar evolution in nearby star-forming dwarf galaxies}},
	volume = {445},
	year = 2014}

@article{chen2014,
	adsurl = {https://ui.adsabs.harvard.edu/abs/2014MNRAS.444.2525C},
	archiveprefix = {arXiv},
	author = {{Chen}, Yang and {Girardi}, L{\'e}o and {Bressan}, Alessandro and {Marigo}, Paola and {Barbieri}, Mauro and {Kong}, Xu},
	doi = {10.1093/mnras/stu1605},
	eprint = {1409.0322},
	journal = {MNRAS},
	month = nov,
	number = {3},
	pages = {2525-2543},
	primaryclass = {astro-ph.SR},
	title = {{Improving PARSEC models for very low mass stars}},
	volume = {444},
	year = 2014}

@article{chen2015,
	adsurl = {https://ui.adsabs.harvard.edu/abs/2015MNRAS.452.1068C},
	archiveprefix = {arXiv},
	author = {{Chen}, Yang and {Bressan}, Alessandro and {Girardi}, L{\'e}o and {Marigo}, Paola and {Kong}, Xu and {Lanza}, Antonio},
	doi = {10.1093/mnras/stv1281},
	eprint = {1506.01681},
	journal = {MNRAS},
	month = sep,
	number = {1},
	pages = {1068-1080},
	primaryclass = {astro-ph.SR},
	title = {{PARSEC evolutionary tracks of massive stars up to 350 M$_{{\ensuremath{\odot}}}$ at metallicities 0.0001 {\ensuremath{\leq}} Z {\ensuremath{\leq}} 0.04}},
	volume = {452},
	year = 2015}

@article{hunt2024,
	adsurl = {https://ui.adsabs.harvard.edu/abs/2024A&A...686A..42H},
	archiveprefix = {arXiv},
	author = {{Hunt}, Emily L. and {Reffert}, Sabine},
	doi = {10.1051/0004-6361/202348662},
	eid = {A42},
	eprint = {2403.05143},
	journal = {A\&A},
	month = jun,
	pages = {A42},
	primaryclass = {astro-ph.GA},
	title = {{Improving the open cluster census. III. Using cluster masses, radii, and dynamics to create a cleaned open cluster catalogue}},
	volume = {686},
	year = 2024}

@article{kounkel2020a,
	adsurl = {https://ui.adsabs.harvard.edu/abs/2020AJ....160..279K},
	archiveprefix = {arXiv},
	author = {{Kounkel}, Marina and {Covey}, Kevin and {Stassun}, Keivan G.},
	doi = {10.3847/1538-3881/abc0e6},
	eid = {279},
	eprint = {2004.07261},
	journal = {AJ},
	month = dec,
	number = {6},
	pages = {279},
	primaryclass = {astro-ph.GA},
	title = {{Untangling the Galaxy. II. Structure within 3 kpc}},
	volume = {160},
	year = 2020}

@article{hunt2023,
	adsurl = {https://ui.adsabs.harvard.edu/abs/2023A&A...673A.114H},
	archiveprefix = {arXiv},
	author = {{Hunt}, Emily L. and {Reffert}, Sabine},
	doi = {10.1051/0004-6361/202346285},
	eid = {A114},
	eprint = {2303.13424},
	journal = {A\&A},
	month = may,
	pages = {A114},
	primaryclass = {astro-ph.GA},
	title = {{Improving the open cluster census. II. An all-sky cluster catalogue with Gaia DR3}},
	volume = {673},
	year = 2023}

@article{mcinnes2017,
	adsurl = {https://ui.adsabs.harvard.edu/abs/2017JOSS....2..205M},
	author = {{McInnes}, Leland and {Healy}, John and {Astels}, Steve},
	doi = {10.21105/joss.00205},
	eid = {205},
	journal = {The Journal of Open Source Software},
	month = mar,
	number = {11},
	pages = {205},
	title = {{hdbscan: Hierarchical density based clustering}},
	volume = {2},
	year = 2017}

@article{larson1981,
	adsurl = {https://ui.adsabs.harvard.edu/abs/1981MNRAS.194..809L},
	author = {{Larson}, R.~B.},
	doi = {10.1093/mnras/194.4.809},
	journal = {MNRAS},
	month = mar,
	pages = {809-826},
	title = {{Turbulence and star formation in molecular clouds.}},
	volume = {194},
	year = 1981}

@article{zhou2022,
	adsurl = {https://ui.adsabs.harvard.edu/abs/2022MNRAS.513..638Z},
	archiveprefix = {arXiv},
	author = {{Zhou}, Ji-Xuan and {Li}, Guang-Xing and {Chen}, Bing-Qiu},
	doi = {10.1093/mnras/stac900},
	eprint = {2110.11595},
	journal = {MNRAS},
	month = jun,
	number = {1},
	pages = {638-647},
	primaryclass = {astro-ph.GA},
	title = {{Kinematics of the molecular interstellar medium probed by Gaia: steep velocity dispersion-size relation, isotropic turbulence, and location-dependent energy dissipation}},
	volume = {513},
	year = 2022}

@article{may1997,
	adsurl = {https://ui.adsabs.harvard.edu/abs/1997A&A...327..325M},
	author = {{May}, J. and {Alvarez}, H. and {Bronfman}, L.},
	journal = {A\&A},
	month = nov,
	pages = {325-332},
	title = {{Physical properties of molecular clouds in the southern outer Galaxy.}},
	volume = {327},
	year = 1997}

@article{thomas2016,
	adsurl = {https://ui.adsabs.harvard.edu/abs/2016ApJ...822...52R},
	archiveprefix = {arXiv},
	author = {{Rice}, Thomas S. and {Goodman}, Alyssa A. and {Bergin}, Edwin A. and {Beaumont}, Christopher and {Dame}, T.~M.},
	doi = {10.3847/0004-637X/822/1/52},
	eid = {52},
	eprint = {1602.02791},
	journal = {ApJ},
	month = may,
	number = {1},
	pages = {52},
	primaryclass = {astro-ph.GA},
	title = {{A Uniform Catalog of Molecular Clouds in the Milky Way}},
	volume = {822},
	year = 2016}

@article{solomon1987,
	adsurl = {https://ui.adsabs.harvard.edu/abs/1987ApJ...319..730S},
	author = {{Solomon}, P.~M. and {Rivolo}, A.~R. and {Barrett}, J. and {Yahil}, A.},
	doi = {10.1086/165493},
	journal = {ApJ},
	month = aug,
	pages = {730},
	title = {{Mass, Luminosity, and Line Width Relations of Galactic Molecular Clouds}},
	volume = {319},
	year = 1987}

@article{caselli1995,
	adsurl = {https://ui.adsabs.harvard.edu/abs/1995ApJ...446..665C},
	author = {{Caselli}, P. and {Myers}, P.~C.},
	doi = {10.1086/175825},
	journal = {ApJ},
	month = jun,
	pages = {665},
	title = {{The Line Width--Size Relation in Massive Cloud Cores}},
	volume = {446},
	year = 1995}

@article{sun2024,
	adsurl = {https://ui.adsabs.harvard.edu/abs/2024ApJS..275...35S},
	author = {{Sun}, Yan and {Yang}, Ji and {Yan}, Qing-Zeng and {Zhang}, Shaobo and {Su}, Yang and {Chen}, Xuepeng and {Zhou}, Xin and {Ma}, Yuehui and {Yuan}, Lixia},
	doi = {10.3847/1538-4365/ad8237},
	eid = {35},
	journal = {ApJS},
	month = dec,
	number = {2},
	pages = {35},
	title = {{Molecular Clouds in the Outer Milky Way Disk: Sample, Integrated Properties, and Radial Trends with Galactocentric Radius}},
	volume = {275},
	year = 2024}

@article{huang2023,
	adsurl = {https://ui.adsabs.harvard.edu/abs/2023ApJ...949...46H},
	archiveprefix = {arXiv},
	author = {{Huang}, Bo and {Wang}, Ke and {Girart}, Josep M. and {Jiao}, Wenyu and {He}, Qianru and {Liang}, Enwei},
	doi = {10.3847/1538-4357/acc532},
	eid = {46},
	eprint = {2303.07501},
	journal = {ApJ},
	month = jun,
	number = {2},
	pages = {46},
	primaryclass = {astro-ph.GA},
	title = {{High-mass Starless Clumps: Dynamical State and Correlation between Physical Parameters}},
	volume = {949},
	year = 2023}

@article{Gaia2016,
	adsurl = {https://ui.adsabs.harvard.edu/abs/2016A&A...595A...1G},
	archiveprefix = {arXiv},
	author = {{Gaia Collaboration} and {Prusti}, T. and {de Bruijne}, J.~H.~J. and {Brown}, A.~G.~A. and {Vallenari}, A. and {Babusiaux}, C. and {Bailer-Jones}, C.~A.~L. and {Bastian}, U. and {Biermann}, M. and {Evans}, D.~W. and {Eyer}, L. and {Jansen}, F. and {Jordi}, C. and {Klioner}, S.~A. and {Lammers}, U. and {Lindegren}, L. and {Luri}, X. and {Mignard}, F. and {Milligan}, D.~J. and {Panem}, C. and {Poinsignon}, V. and {Pourbaix}, D. and {Randich}, S. and {Sarri}, G. and {Sartoretti}, P. and {Siddiqui}, H.~I. and {Soubiran}, C. and {Valette}, V. and {van Leeuwen}, F. and {Walton}, N.~A. and {Aerts}, C. and {Arenou}, F. and {Cropper}, M. and {Drimmel}, R. and {H{\o}g}, E. and {Katz}, D. and {Lattanzi}, M.~G. and {O'Mullane}, W. and {Grebel}, E.~K. and {Holland}, A.~D. and {Huc}, C. and {Passot}, X. and {Bramante}, L. and {Cacciari}, C. and {Casta{\~n}eda}, J. and {Chaoul}, L. and {Cheek}, N. and {De Angeli}, F. and {Fabricius}, C. and {Guerra}, R. and {Hern{\'a}ndez}, J. and {Jean-Antoine-Piccolo}, A. and {Masana}, E. and {Messineo}, R. and {Mowlavi}, N. and {Nienartowicz}, K. and {Ord{\'o}{\~n}ez-Blanco}, D. and {Panuzzo}, P. and {Portell}, J. and {Richards}, P.~J. and {Riello}, M. and {Seabroke}, G.~M. and {Tanga}, P. and {Th{\'e}venin}, F. and {Torra}, J. and {Els}, S.~G. and {Gracia-Abril}, G. and {Comoretto}, G. and {Garcia-Reinaldos}, M. and {Lock}, T. and {Mercier}, E. and {Altmann}, M. and {Andrae}, R. and {Astraatmadja}, T.~L. and {Bellas-Velidis}, I. and {Benson}, K. and {Berthier}, J. and {Blomme}, R. and {Busso}, G. and {Carry}, B. and {Cellino}, A. and {Clementini}, G. and {Cowell}, S. and {Creevey}, O. and {Cuypers}, J. and {Davidson}, M. and {De Ridder}, J. and {de Torres}, A. and {Delchambre}, L. and {Dell'Oro}, A. and {Ducourant}, C. and {Fr{\'e}mat}, Y. and {Garc{\'\i}a-Torres}, M. and {Gosset}, E. and {Halbwachs}, J. -L. and {Hambly}, N.~C. and {Harrison}, D.~L. and {Hauser}, M. and {Hestroffer}, D. and {Hodgkin}, S.~T. and {Huckle}, H.~E. and {Hutton}, A. and {Jasniewicz}, G. and {Jordan}, S. and {Kontizas}, M. and {Korn}, A.~J. and {Lanzafame}, A.~C. and {Manteiga}, M. and {Moitinho}, A. and {Muinonen}, K. and {Osinde}, J. and {Pancino}, E. and {Pauwels}, T. and {Petit}, J. -M. and {Recio-Blanco}, A. and {Robin}, A.~C. and {Sarro}, L.~M. and {Siopis}, C. and {Smith}, M. and {Smith}, K.~W. and {Sozzetti}, A. and {Thuillot}, W. and {van Reeven}, W. and {Viala}, Y. and {Abbas}, U. and {Abreu Aramburu}, A. and {Accart}, S. and {Aguado}, J.~J. and {Allan}, P.~M. and {Allasia}, W. and {Altavilla}, G. and {{\'A}lvarez}, M.~A. and {Alves}, J. and {Anderson}, R.~I. and {Andrei}, A.~H. and {Anglada Varela}, E. and {Antiche}, E. and {Antoja}, T. and {Ant{\'o}n}, S. and {Arcay}, B. and {Atzei}, A. and {Ayache}, L. and {Bach}, N. and {Baker}, S.~G. and {Balaguer-N{\'u}{\~n}ez}, L. and {Barache}, C. and {Barata}, C. and {Barbier}, A. and {Barblan}, F. and {Baroni}, M. and {Barrado y Navascu{\'e}s}, D. and {Barros}, M. and {Barstow}, M.~A. and {Becciani}, U. and {Bellazzini}, M. and {Bellei}, G. and {Bello Garc{\'\i}a}, A. and {Belokurov}, V. and {Bendjoya}, P. and {Berihuete}, A. and {Bianchi}, L. and {Bienaym{\'e}}, O. and {Billebaud}, F. and {Blagorodnova}, N. and {Blanco-Cuaresma}, S. and {Boch}, T. and {Bombrun}, A. and {Borrachero}, R. and {Bouquillon}, S. and {Bourda}, G. and {Bouy}, H. and {Bragaglia}, A. and {Breddels}, M.~A. and {Brouillet}, N. and {Br{\"u}semeister}, T. and {Bucciarelli}, B. and {Budnik}, F. and {Burgess}, P. and {Burgon}, R. and {Burlacu}, A. and {Busonero}, D. and {Buzzi}, R. and {Caffau}, E. and {Cambras}, J. and {Campbell}, H. and {Cancelliere}, R. and {Cantat-Gaudin}, T. and {Carlucci}, T. and {Carrasco}, J.~M. and {Castellani}, M. and {Charlot}, P. and {Charnas}, J. and {Charvet}, P. and {Chassat}, F. and {Chiavassa}, A. and {Clotet}, M. and {Cocozza}, G. and {Collins}, R.~S. and {Collins}, P. and {Costigan}, G.},
	doi = {10.1051/0004-6361/201629272},
	eid = {A1},
	eprint = {1609.04153},
	journal = {A\&A},
	month = nov,
	pages = {A1},
	primaryclass = {astro-ph.IM},
	title = {{The Gaia mission}},
	volume = {595},
	year = 2016}

@article{li2022,
	adsurl = {https://ui.adsabs.harvard.edu/abs/2022MNRAS.517L.102L},
	archiveprefix = {arXiv},
	author = {{Li}, Guang-Xing and {Chen}, Bing-Qiu},
	doi = {10.1093/mnrasl/slac050},
	eprint = {2205.03218},
	journal = {MNRAS},
	month = nov,
	number = {1},
	pages = {L102-L107},
	primaryclass = {astro-ph.GA},
	title = {{Discovery of a coherent, wave-like velocity pattern for the Radcliffe wave}},
	volume = {517},
	year = 2022}

@article{lim2021,
	adsurl = {https://ui.adsabs.harvard.edu/abs/2021PASJ...73S.239L},
	archiveprefix = {arXiv},
	author = {{Lim}, Wanggi and {Nakamura}, Fumitaka and {Wu}, Benjamin and {Bisbas}, Thomas G. and {Tan}, Jonathan C. and {Chambers}, Edward and {Bally}, John and {Kong}, Shuo and {McGehee}, Peregrine and {Lis}, Dariusz C. and {Ossenkopf-Okada}, Volker and {S{\'a}nchez-Monge}, {\'A}lvaro},
	doi = {10.1093/pasj/psaa035},
	eprint = {2004.03668},
	journal = {PASJ},
	month = jan,
	pages = {S239-S255},
	primaryclass = {astro-ph.SR},
	title = {{Star cluster formation in Orion A}},
	volume = {73},
	year = 2021}

@article{ha2021,
	adsurl = {https://ui.adsabs.harvard.edu/abs/2021ApJ...907L..40H},
	archiveprefix = {arXiv},
	author = {{Ha}, Trung and {Li}, Yuan and {Xu}, Siyao and {Kounkel}, Marina and {Li}, Hui},
	doi = {10.3847/2041-8213/abd8c9},
	eid = {L40},
	eprint = {2101.03176},
	journal = {ApJL},
	month = feb,
	number = {2},
	pages = {L40},
	primaryclass = {astro-ph.GA},
	title = {{Measuring Turbulence with Young Stars in the Orion Complex}},
	volume = {907},
	year = 2021}

@article{zhou2024,
	adsurl = {https://ui.adsabs.harvard.edu/abs/2024MNRAS.529.1091Z},
	archiveprefix = {arXiv},
	author = {{Zhou}, Ji-Xuan and {Li}, Guang-Xing and {Chen}, Bing-Qiu},
	doi = {10.1093/mnras/stae376},
	eprint = {2402.02393},
	journal = {MNRAS},
	month = apr,
	number = {2},
	pages = {1091-1103},
	primaryclass = {astro-ph.GA},
	title = {{Gas content and evolution of a sample of YSO associations at d {\ensuremath{\lesssim}} 3.5 kpc from the Sun}},
	volume = {529},
	year = 2024}

@article{perrot2003,
	adsurl = {https://ui.adsabs.harvard.edu/abs/2003A&A...404..519P},
	archiveprefix = {arXiv},
	author = {{Perrot}, C.~A. and {Grenier}, I.~A.},
	doi = {10.1051/0004-6361:20030477},
	eprint = {astro-ph/0303516},
	journal = {A\&A},
	month = jun,
	pages = {519-531},
	primaryclass = {astro-ph},
	title = {{3D dynamical evolution of the interstellar gas in the Gould Belt}},
	volume = {404},
	year = 2003}

@article{lallement2019,
	adsurl = {https://ui.adsabs.harvard.edu/abs/2019A&A...625A.135L},
	archiveprefix = {arXiv},
	author = {{Lallement}, R. and {Babusiaux}, C. and {Vergely}, J.~L. and {Katz}, D. and {Arenou}, F. and {Valette}, B. and {Hottier}, C. and {Capitanio}, L.},
	doi = {10.1051/0004-6361/201834695},
	eid = {A135},
	eprint = {1902.04116},
	journal = {A\&A},
	month = may,
	pages = {A135},
	primaryclass = {astro-ph.GA},
	title = {{Gaia-2MASS 3D maps of Galactic interstellar dust within 3 kpc}},
	volume = {625},
	year = 2019}

@article{alves2020,
	adsurl = {https://ui.adsabs.harvard.edu/abs/2020Natur.578..237A},
	archiveprefix = {arXiv},
	author = {{Alves}, Jo{\~a}o and {Zucker}, Catherine and {Goodman}, Alyssa A. and {Speagle}, Joshua S. and {Meingast}, Stefan and {Robitaille}, Thomas and {Finkbeiner}, Douglas P. and {Schlafly}, Edward F. and {Green}, Gregory M.},
	doi = {10.1038/s41586-019-1874-z},
	eprint = {2001.08748},
	journal = {Nature},
	month = feb,
	number = {7794},
	pages = {237-239},
	primaryclass = {astro-ph.GA},
	title = {{A Galactic-scale gas wave in the solar neighbourhood}},
	volume = {578},
	year = 2020}

@article{pm2016,
	adsurl = {https://ui.adsabs.harvard.edu/abs/2016MNRAS.461..794P},
	archiveprefix = {arXiv},
	author = {{Pecaut}, Mark J. and {Mamajek}, Eric E.},
	doi = {10.1093/mnras/stw1300},
	eprint = {1605.08789},
	journal = {MNRAS},
	month = sep,
	number = {1},
	pages = {794-815},
	primaryclass = {astro-ph.SR},
	title = {{The star formation history and accretion-disc fraction among the K-type members of the Scorpius-Centaurus OB association}},
	volume = {461},
	year = 2016}

@ARTICLE{bensby2014,
       author = {{Bensby}, T. and {Feltzing}, S. and {Oey}, M.~S.},
        title = "{Exploring the Milky Way stellar disk. A detailed elemental abundance study of 714 F and G dwarf stars in the solar neighbourhood}",
      journal = {A\&A},
         year = 2014,
        month = feb,
       volume = {562},
          eid = {A71},
        pages = {A71},
          doi = {10.1051/0004-6361/201322631},
archivePrefix = {arXiv},
       eprint = {1309.2631},
 primaryClass = {astro-ph.GA},
       adsurl = {https://ui.adsabs.harvard.edu/abs/2014A&A...562A..71B}
}

@ARTICLE{fleck1980,
       author = {{Fleck}, Jr., R.~C.},
        title = "{Turbulence and the stability of molecular clouds.}",
      journal = {\apj},
         year = 1980,
        month = dec,
       volume = {242},
        pages = {1019-1022},
          doi = {10.1086/158533},
       adsurl = {https://ui.adsabs.harvard.edu/abs/1980ApJ...242.1019F}
}

@ARTICLE{mac2004,
       author = {{Mac Low}, Mordecai-Mark and {Klessen}, Ralf S.},
        title = "{Control of star formation by supersonic turbulence}",
      journal = {Reviews of Modern Physics},
         year = 2004,
        month = jan,
       volume = {76},
       number = {1},
        pages = {125-194},
          doi = {10.1103/RevModPhys.76.125},
archivePrefix = {arXiv},
       eprint = {astro-ph/0301093},
 primaryClass = {astro-ph},
       adsurl = {https://ui.adsabs.harvard.edu/abs/2004RvMP...76..125M}
}

@ARTICLE{elme2004,
       author = {{Elmegreen}, Bruce G. and {Scalo}, John},
        title = "{Interstellar Turbulence I: Observations and Processes}",
      journal = {\araa},
         year = 2004,
        month = sep,
       volume = {42},
       number = {1},
        pages = {211-273},
          doi = {10.1146/annurev.astro.41.011802.094859},
archivePrefix = {arXiv},
       eprint = {astro-ph/0404451},
 primaryClass = {astro-ph},
       adsurl = {https://ui.adsabs.harvard.edu/abs/2004ARA&A..42..211E}
}

@ARTICLE{balle2011,
       author = {{Ballesteros-Paredes}, Javier and {Hartmann}, Lee W. and {V{\'a}zquez-Semadeni}, Enrique and {Heitsch}, Fabian and {Zamora-Avil{\'e}s}, Manuel A.},
        title = "{Gravity or turbulence? Velocity dispersion-size relation}",
      journal = {\mnras},
         year = 2011,
        month = feb,
       volume = {411},
       number = {1},
        pages = {65-70},
          doi = {10.1111/j.1365-2966.2010.17657.x},
archivePrefix = {arXiv},
       eprint = {1009.1583},
 primaryClass = {astro-ph.GA},
       adsurl = {https://ui.adsabs.harvard.edu/abs/2011MNRAS.411...65B}
}

@ARTICLE{izqui2021,
       author = {{Izquierdo}, Andr{\'e}s F. and {Smith}, Rowan J. and {Glover}, Simon C.~O. and {Klessen}, Ralf S. and {Tre{\ss}}, Robin G. and {Sormani}, Mattia C. and {Clark}, Paul C. and {Duarte-Cabral}, Ana and {Zucker}, Catherine},
        title = "{The Cloud Factory II: gravoturbulent kinematics of resolved molecular clouds in a galactic potential}",
      journal = {\mnras},
         year = 2021,
        month = jan,
       volume = {500},
       number = {4},
        pages = {5268-5296},
          doi = {10.1093/mnras/staa3470},
archivePrefix = {arXiv},
       eprint = {2011.02582},
 primaryClass = {astro-ph.GA},
       adsurl = {https://ui.adsabs.harvard.edu/abs/2021MNRAS.500.5268I}
}

@ARTICLE{xie2025,
       author = {{Xie}, Yi-Heng and {Li}, Guang-Xing},
        title = "{Shear{\textendash}Gravity Transition Determines the Steep Velocity Dispersion{\textendash}Size Relation in Molecular Clouds: Confronting Analytical Formula with Observations}",
      journal = {\apjl},
         year = 2025,
        month = apr,
       volume = {983},
       number = {1},
          eid = {L21},
        pages = {L21},
          doi = {10.3847/2041-8213/adc095},
archivePrefix = {arXiv},
       eprint = {2501.03027},
 primaryClass = {astro-ph.GA},
       adsurl = {https://ui.adsabs.harvard.edu/abs/2025ApJ...983L..21X}
}

@ARTICLE{jackson2006,
       author = {{Jackson}, J.~M. and {Rathborne}, J.~M. and {Shah}, R.~Y. and {Simon}, R. and {Bania}, T.~M. and {Clemens}, D.~P. and {Chambers}, E.~T. and {Johnson}, A.~M. and {Dormody}, M. and {Lavoie}, R. and {Heyer}, M.~H.},
        title = "{The Boston University-Five College Radio Astronomy Observatory Galactic Ring Survey}",
      journal = {\apjs},
         year = 2006,
        month = mar,
       volume = {163},
       number = {1},
        pages = {145-159},
          doi = {10.1086/500091},
archivePrefix = {arXiv},
       eprint = {astro-ph/0602160},
 primaryClass = {astro-ph},
       adsurl = {https://ui.adsabs.harvard.edu/abs/2006ApJS..163..145J}
}

@ARTICLE{ligx2024,
       author = {{Li}, Guang-Xing},
        title = "{Tides in clouds: control of star formation by long-range gravitational force}",
      journal = {\mnras},
         year = 2024,
        month = feb,
       volume = {528},
       number = {1},
        pages = {L52-L58},
          doi = {10.1093/mnrasl/slad149},
archivePrefix = {arXiv},
       eprint = {2309.16125},
 primaryClass = {astro-ph.GA},
       adsurl = {https://ui.adsabs.harvard.edu/abs/2024MNRAS.528L..52L}
}

@ARTICLE{crut2012,
       author = {{Crutcher}, Richard M.},
        title = "{Magnetic Fields in Molecular Clouds}",
      journal = {\araa},
         year = 2012,
        month = sep,
       volume = {50},
        pages = {29-63},
          doi = {10.1146/annurev-astro-081811-125514},
       adsurl = {https://ui.adsabs.harvard.edu/abs/2012ARA&A..50...29C}
}

@ARTICLE{ligx2022,
       author = {{Li}, Guang-Xing and {Zhou}, Ji-Xuan and {Chen}, Bing-Qiu},
        title = "{Weather forecast of the Milky Way: shear and stellar feedback determine the lives of Galactic-scale filaments}",
      journal = {\mnras},
         year = 2022,
        month = oct,
       volume = {516},
       number = {1},
        pages = {L35-L42},
          doi = {10.1093/mnrasl/slac076},
archivePrefix = {arXiv},
       eprint = {2207.03835},
 primaryClass = {astro-ph.GA},
       adsurl = {https://ui.adsabs.harvard.edu/abs/2022MNRAS.516L..35L}
}

@ARTICLE{kolmo1941,
       author = {{Kolmogorov}, A.},
        title = "{The Local Structure of Turbulence in Incompressible Viscous Fluid for Very Large Reynolds' Numbers}",
      journal = {Akademiia Nauk SSSR Doklady},
         year = 1941,
        month = jan,
       volume = {30},
        pages = {301-305},
       adsurl = {https://ui.adsabs.harvard.edu/abs/1941DoSSR..30..301K}
}

@ARTICLE{Zhang2025,
       author = {{Zhang}, Xiangyu and {Green}, Gregory M.},
        title = "{Three-dimensional maps of the interstellar dust extinction curve within the Milky Way galaxy}",
      journal = {Science},
         year = 2025,
        month = mar,
       volume = {387},
       number = {6739},
        pages = {1209-1214},
          doi = {10.1126/science.ado9787},
archivePrefix = {arXiv},
       eprint = {2407.14594},
 primaryClass = {astro-ph.GA},
       adsurl = {https://ui.adsabs.harvard.edu/abs/2025Sci...387.1209Z}
}

@INPROCEEDINGS{Ma2005,
       author = {{Ma{\'\i}z Apell{\'a}niz}, J.},
        title = "{Self-Consistent Distance Determinations for Lutz-Kelker-Limited Samples}",
    booktitle = {The Three-Dimensional Universe with Gaia},
         year = 2005,
       editor = {{Turon}, C. and {O'Flaherty}, K.~S. and {Perryman}, M.~A.~C.},
       series = {ESA Special Publication},
       volume = {576},
        month = jan,
        pages = {179},
          doi = {10.48550/arXiv.astro-ph/0411346},
archivePrefix = {arXiv},
       eprint = {astro-ph/0411346},
 primaryClass = {astro-ph},
       adsurl = {https://ui.adsabs.harvard.edu/abs/2005ESASP.576..179M}
}

@ARTICLE{Chen2025,
       author = {{Chen}, Bingqiu and {Li}, Guangxing and {Yuan}, Haibo and {Xiang}, Maosheng and {Zhou}, Jixuan and {Chen}, Pinjian and {Krause}, Martin and {Coombs}, Ashley},
        title = "{A large, long-lived, slowly-expanding superbubble across the Perseus arm}",
      journal = {Nature Communications},
         year = 2025,
        month = nov,
       volume = {16},
       number = {1},
          eid = {10558},
        pages = {10558},
          doi = {10.1038/s41467-025-65591-5},
archivePrefix = {arXiv},
       eprint = {2512.21927},
 primaryClass = {astro-ph.GA},
       adsurl = {https://ui.adsabs.harvard.edu/abs/2025NatCo..1610558C}
}

@INPROCEEDINGS{Zucker2023,
       author = {{Zucker}, C. and {Alves}, J. and {Goodman}, A. and {Meingast}, S. and {Galli}, P.},
        title = "{The Solar Neighborhood in the Age of Gaia}",
    booktitle = {Protostars and Planets VII},
         year = 2023,
       editor = {{Inutsuka}, S. and {Aikawa}, Y. and {Muto}, T. and {Tomida}, K. and {Tamura}, M.},
       series = {Astronomical Society of the Pacific Conference Series},
       volume = {534},
        month = jul,
        pages = {43},
          doi = {10.48550/arXiv.2212.00067},
archivePrefix = {arXiv},
       eprint = {2212.00067},
 primaryClass = {astro-ph.GA},
       adsurl = {https://ui.adsabs.harvard.edu/abs/2023ASPC..534...43Z}
}

@INPROCEEDINGS{Lada1987,
       author = {{Lada}, Charles J.},
        title = "{Star formation: from OB associations to protostars.}",
    booktitle = {Star Forming Regions},
         year = 1987,
       editor = {{Peimbert}, Manuel and {Jugaku}, Jun},
       series = {IAU Symposium},
       volume = {115},
        month = jan,
        pages = {1},
       adsurl = {https://ui.adsabs.harvard.edu/abs/1987IAUS..115....1L}
}

@ARTICLE{Andre1993,
       author = {{Andre}, Philippe and {Ward-Thompson}, Derek and {Barsony}, Mary},
        title = "{Submillimeter Continuum Observations of rho Ophiuchi A: The Candidate Protostar VLA 1623 and Prestellar Clumps}",
      journal = {\apj},
         year = 1993,
        month = mar,
       volume = {406},
        pages = {122},
          doi = {10.1086/172425},
       adsurl = {https://ui.adsabs.harvard.edu/abs/1993ApJ...406..122A}
}

@ARTICLE{Greene1994,
       author = {{Greene}, Thomas P. and {Wilking}, Bruce A. and {Andre}, Philippe and {Young}, Erick T. and {Lada}, Charles J.},
        title = "{Further Mid-Infrared Study of the rho Ophiuchi Cloud Young Stellar Population: Luminosities and Masses of Pre--Main-Sequence Stars}",
      journal = {\apj},
         year = 1994,
        month = oct,
       volume = {434},
        pages = {614},
          doi = {10.1086/174763},
       adsurl = {https://ui.adsabs.harvard.edu/abs/1994ApJ...434..614G}
}

@ARTICLE{Allen2004,
       author = {{Allen}, Lori E. and {Calvet}, Nuria and {D'Alessio}, Paola and {Merin}, Bruno and {Hartmann}, Lee and {Megeath}, S. Thomas and {Gutermuth}, Robert A. and {Muzerolle}, James and {Pipher}, Judith L. and {Myers}, Philip C. and {Fazio}, Giovanni G.},
        title = "{Infrared Array Camera (IRAC) Colors of Young Stellar Objects}",
      journal = {\apjs},
         year = 2004,
        month = sep,
       volume = {154},
       number = {1},
        pages = {363-366},
          doi = {10.1086/422715},
       adsurl = {https://ui.adsabs.harvard.edu/abs/2004ApJS..154..363A}
}

@INPROCEEDINGS{Dunham2014,
       author = {{Dunham}, M.~M. and {Stutz}, A.~M. and {Allen}, L.~E. and {Evans}, II, N.~J. and {Fischer}, W.~J. and {Megeath}, S.~T. and {Myers}, P.~C. and {Offner}, S.~S.~R. and {Poteet}, C.~A. and {Tobin}, J.~J. and et al.},
        title = "{The Evolution of Protostars: Insights from Ten Years of Infrared Surveys with Spitzer and Herschel}",
    booktitle = {Protostars and Planets VI},
         year = 2014,
       editor = {{Beuther}, Henrik and {Klessen}, Ralf S. and {Dullemond}, Cornelis P. and {Henning}, Thomas},
        month = jan,
        pages = {195-218},
          doi = {10.2458/azu_uapress_9780816531240-ch009},
archivePrefix = {arXiv},
       eprint = {1401.1809},
 primaryClass = {astro-ph.GA},
       adsurl = {https://ui.adsabs.harvard.edu/abs/2014prpl.conf..195D}
}

@ARTICLE{Ma2024,
       author = {{Ma}, Xiangyao and {Zhang}, Yanxia and {Zhang}, Jingyi and {Li}, Changhua and {Kang}, Zihan and {Li}, Ji},
        title = "{Search for Young Stellar Objects within 4XMM-DR13 Using CatBoost and SPE}",
      journal = {AJ},
         year = 2024,
        month = nov,
       volume = {168},
       number = {5},
          eid = {210},
        pages = {210},
          doi = {10.3847/1538-3881/ad781c},
archivePrefix = {arXiv},
       eprint = {2410.11436},
 primaryClass = {astro-ph.GA},
       adsurl = {https://ui.adsabs.harvard.edu/abs/2024AJ....168..210M}
}

@ARTICLE{Marton2023,
       author = {{Marton}, G{\'a}bor and {{\'A}brah{\'a}m}, P{\'e}ter and {Rimoldini}, Lorenzo and {Audard}, Marc and {Kun}, M{\'a}ria and {Nagy}, Zs{\'o}fia and {K{\'o}sp{\'a}l}, {\'A}gnes and {Szabados}, L{\'a}szl{\'o} and {Holl}, Berry and {Gavras}, Panagiotis and et al.},
        title = "{Gaia Data Release 3. Validating the classification of variable young stellar object candidates}",
      journal = {A\&A},
         year = 2023,
        month = jun,
       volume = {674},
          eid = {A21},
        pages = {A21},
          doi = {10.1051/0004-6361/202244101},
archivePrefix = {arXiv},
       eprint = {2206.05796},
 primaryClass = {astro-ph.SR},
       adsurl = {https://ui.adsabs.harvard.edu/abs/2023A&A...674A..21M}
}

@ARTICLE{Rimoldini2023,
       author = {{Rimoldini}, Lorenzo and {Holl}, Berry and {Gavras}, Panagiotis and {Audard}, Marc and {De Ridder}, Joris and {Mowlavi}, Nami and {Nienartowicz}, Krzysztof and {Jevardat de Fombelle}, Gr{\'e}gory and {Lecoeur-Ta{\"\i}bi}, Isabelle and {Karbevska}, Lea and et al.},
        title = "{Gaia Data Release 3. All-sky classification of 12.4 million variable sources into 25 classes}",
      journal = {A\&A},
         year = 2023,
        month = jun,
       volume = {674},
          eid = {A14},
        pages = {A14},
          doi = {10.1051/0004-6361/202245591},
archivePrefix = {arXiv},
       eprint = {2211.17238},
 primaryClass = {astro-ph.GA},
       adsurl = {https://ui.adsabs.harvard.edu/abs/2023A&A...674A..14R}
}

@ARTICLE{Kerr2023,
       author = {{Kerr}, Ronan and {Kraus}, Adam L. and {Rizzuto}, Aaron C.},
        title = "{SPYGLASS. IV. New Stellar Survey of Recent Star Formation within 1 kpc}",
      journal = {\apj},
         year = 2023,
        month = sep,
       volume = {954},
       number = {2},
          eid = {134},
        pages = {134},
          doi = {10.3847/1538-4357/ace5b3},
archivePrefix = {arXiv},
       eprint = {2306.08150},
 primaryClass = {astro-ph.GA},
       adsurl = {https://ui.adsabs.harvard.edu/abs/2023ApJ...954..134K}
}

@ARTICLE{Quintana2026,
       author = {{Quintana}, Alexis L. and {Wright}, Nicholas J. and {Kormann}, Lilly A. and {Alves}, Jo{\~a}o and {Katz}, David and {Casamiquela}, Laia and {Di Matteo}, Paola and {Haywood}, Misha and {Laporte}, Chervin},
        title = "{A new Gaia census of OB associations within 1 kpc}",
      journal = {\mnras},
         year = 2026,
        month = jun,
       volume = {549},
       number = {1},
          eid = {stag853},
        pages = {stag853},
          doi = {10.1093/mnras/stag853},
archivePrefix = {arXiv},
       eprint = {2512.05854},
 primaryClass = {astro-ph.GA},
       adsurl = {https://ui.adsabs.harvard.edu/abs/2026MNRAS.549ag853Q}
}

@ARTICLE{Kormann2026,
       author = {{Kormann}, Lilly A. and {Alves}, Jo{\~a}o and {Pantaleoni Gonz{\'a}lez}, Michelangelo and {Swiggum}, Cameren and {En{\ss}lin}, Torsten A. and {Edenhofer}, Gordian},
        title = "{The superclouds of the local Milky Way}",
      journal = {A\&A},
         year = 2026,
        month = feb,
       volume = {706},
          eid = {A161},
        pages = {A161},
          doi = {10.1051/0004-6361/202556469},
archivePrefix = {arXiv},
       eprint = {2507.14883},
 primaryClass = {astro-ph.GA},
       adsurl = {https://ui.adsabs.harvard.edu/abs/2026A&A...706A.161K}
}

@ARTICLE{Miville2017,
       author = {{Miville-Desch{\^e}nes}, Marc-Antoine and {Murray}, Norman and {Lee}, Eve J.},
        title = "{Physical Properties of Molecular Clouds for the Entire Milky Way Disk}",
      journal = {\apj},
         year = 2017,
        month = jan,
       volume = {834},
       number = {1},
          eid = {57},
        pages = {57},
          doi = {10.3847/1538-4357/834/1/57},
archivePrefix = {arXiv},
       eprint = {1610.05918},
 primaryClass = {astro-ph.GA},
       adsurl = {https://ui.adsabs.harvard.edu/abs/2017ApJ...834...57M}
}

@ARTICLE{Quintana2022,
       author = {{Quintana}, Alexis L. and {Wright}, Nicholas J.},
        title = "{Large-scale expansion of OB stars in Cygnus}",
      journal = {\mnras},
         year = 2022,
        month = sep,
       volume = {515},
       number = {1},
        pages = {687-692},
          doi = {10.1093/mnras/stac1526},
archivePrefix = {arXiv},
       eprint = {2205.15611},
 primaryClass = {astro-ph.SR},
       adsurl = {https://ui.adsabs.harvard.edu/abs/2022MNRAS.515..687Q}
}

@ARTICLE{Kounkel2020b,
       author = {{Kounkel}, Marina},
        title = "{Supernovae in Orion: The Missing Link in the Star-forming History of the Region}",
      journal = {\apj},
         year = 2020,
        month = oct,
       volume = {902},
       number = {2},
          eid = {122},
        pages = {122},
          doi = {10.3847/1538-4357/abb6e8},
archivePrefix = {arXiv},
       eprint = {2007.09160},
 primaryClass = {astro-ph.SR},
       adsurl = {https://ui.adsabs.harvard.edu/abs/2020ApJ...902..122K}
}

@ARTICLE{Drew2021,
       author = {{Drew}, J.~E. and {Mongui{\'o}}, M. and {Wright}, N.~J.},
        title = "{Proper motions of OB stars in the far Carina Arm}",
      journal = {\mnras},
         year = 2021,
        month = dec,
       volume = {508},
       number = {4},
        pages = {4952-4968},
          doi = {10.1093/mnras/stab2905},
archivePrefix = {arXiv},
       eprint = {2110.02081},
 primaryClass = {astro-ph.GA},
       adsurl = {https://ui.adsabs.harvard.edu/abs/2021MNRAS.508.4952D}
}

@ARTICLE{Grossschedl2021,
       author = {{Gro{\ss}schedl}, Josefa E. and {Alves}, Jo{\~a}o and {Meingast}, Stefan and {Herbst-Kiss}, Gabor},
        title = "{3D dynamics of the Orion cloud complex. Discovery of coherent radial gas motions at the 100-pc scale}",
      journal = {A\&A},
         year = 2021,
        month = mar,
       volume = {647},
          eid = {A91},
        pages = {A91},
          doi = {10.1051/0004-6361/202038913},
archivePrefix = {arXiv},
       eprint = {2007.07254},
 primaryClass = {astro-ph.SR},
       adsurl = {https://ui.adsabs.harvard.edu/abs/2021A&A...647A..91G}
}

@ARTICLE{Grossschedl2026,
       author = {{Gro{\ss}schedl}, Josefa E. and {Alves}, Jo{\~a}o and {Ratzenb{\"o}ck}, Sebastian and {Miret-Roig}, N{\'u}ria and {Hutschenreuter}, Sebastian and {Posch}, Laura and {Hacar}, Alvaro},
        title = "{The evolution of velocity dispersion in the Sco-Cen OB association}",
      journal = {A\&A},
         year = 2026,
        month = may,
       volume = {709},
          eid = {A181},
        pages = {A181},
          doi = {10.1051/0004-6361/202555519},
archivePrefix = {arXiv},
       eprint = {2509.19487},
 primaryClass = {astro-ph.GA},
       adsurl = {https://ui.adsabs.harvard.edu/abs/2026A&A...709A.181G}
}

@ARTICLE{Anderson2025,
       author = {{Anderson}, Richard I. and {Hunt}, Emily L.},
        title = "{A bird's eye view of stellar evolution through populations of variable stars in Galactic open clusters}",
      journal = {A\&A},
         year = 2025,
        month = aug,
       volume = {700},
          eid = {L13},
        pages = {L13},
          doi = {10.1051/0004-6361/202555111},
archivePrefix = {arXiv},
       eprint = {2508.12866},
 primaryClass = {astro-ph.SR},
       adsurl = {https://ui.adsabs.harvard.edu/abs/2025A&A...700L..13A}
}

@ARTICLE{Katz2023,
       author = {{Katz}, D. and {Sartoretti}, P. and {Guerrier}, A. and {Panuzzo}, P. and {Seabroke}, G.~M. and {Th{\'e}venin}, F. and {Cropper}, M. and {Benson}, K. and {Blomme}, R. and {Haigron}, R. and et al.},
        title = "{Gaia Data Release 3. Properties and validation of the radial velocities}",
      journal = {\aap},
         year = 2023,
        month = jun,
       volume = {674},
          eid = {A5},
        pages = {A5},
          doi = {10.1051/0004-6361/202244220},
archivePrefix = {arXiv},
       eprint = {2206.05902},
 primaryClass = {astro-ph.GA},
       adsurl = {https://ui.adsabs.harvard.edu/abs/2023A&A...674A...5K}
}

@ARTICLE{Chen2020,
       author = {{Chen}, B.-Q. and {Li}, G.-X. and {Yuan}, H.-B. and {Huang}, Y. and {Tian}, Z.-J. and {Wang}, H.-F. and {Zhang}, H.-W. and {Wang}, C. and {Liu}, X.-W.},
        title = "{A large catalogue of molecular clouds with accurate distances within 4 kpc of the Galactic disc}",
      journal = {\mnras},
         year = 2020,
        month = mar,
       volume = {493},
       number = {1},
        pages = {351-361},
          doi = {10.1093/mnras/staa235},
archivePrefix = {arXiv},
       eprint = {2001.11682},
 primaryClass = {astro-ph.GA},
       adsurl = {https://ui.adsabs.harvard.edu/abs/2020MNRAS.493..351C}
}



\clearpage
\appendix

\section{Comparison with Published Young-Star Samples}
\label{sec:comparison_with_published_young_star_samples}

\begin{table*}
	\centering
	\caption{Cross-matching results between our samples and published young-star catalogues.}
	\label{tab:appendix_crossmatch}
	\begin{tabular}{llrrr}
		\hline
		Reference sample & Sample type & $N_{\rm ref}$\textsuperscript{a} & Initial YSO candidates\textsuperscript{b} & Final member stars\textsuperscript{b}\\
		\hline
		Gaia DR3 variable YSOs & Variable YSO candidates & 79\,375 & 4812 (0.65\%) & 646 (11.31\%)\\
		KYSO catalogue & Optical YSO catalogue & 11\,671 & 1429 (0.19\%) & 348 (6.09\%)\\
		Zari et al.\ (2018) PMS sample & PMS candidates & 43\,719 & 15\,780 (2.13\%) & 1597 (27.95\%) \\
		Kerr et al.\ (2023) SPYGLASS IV & Young stellar associations & 1\,222\,152 & 19\,854 (2.68\%) & 2646 (46.32\%)\\
		\hline
	\end{tabular}
	\begin{flushleft}
	\textbf{Notes.}\\
	\textsuperscript{a} Total number of sources in the reference catalogue.\\
	\textsuperscript{b} Number of matched sources with our sample; the fraction in parentheses is the number of matched sources divided by the total number of sources in our corresponding sample (initial YSO candidates: $\sim$740\,000; final association members: 5713).\\
	\end{flushleft}
\end{table*}

\begin{figure*}
	\centering
	\includegraphics[width=\textwidth]{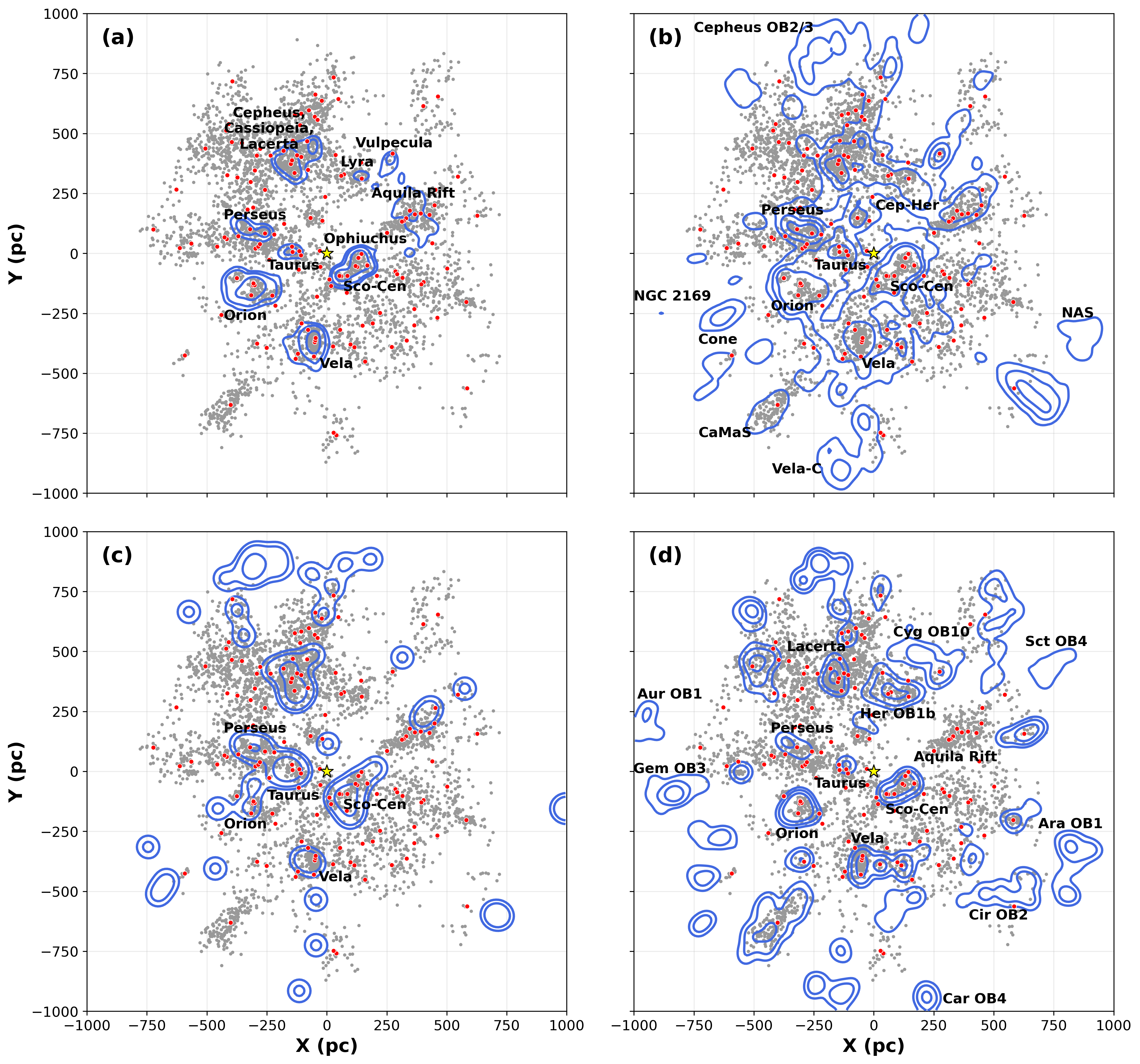}
	\caption{Spatial comparison between our YSO association catalogue and published young-star samples in the Galactic $X$--$Y$ plane. In all panels, red filled circles mark the centres of our YSO groups, grey dots show the positions of our member stars, and the blue contours represent the density distributions of the reference samples. The four panels correspond to: (a) \citet{zari2018} PMS sample; (b) \citet{Kerr2023} SPYGLASS IV young association members; (c) \citet{kounkel2020a} young stellar groups (restricted to distances $<$\,1~kpc and ages $<$\,20~Myr); (d) \citet{Quintana2026} OB association members. The yellow star marks the position of the Sun. The names of selected stellar associations and star-forming regions discussed in the text are labelled.}
	\label{fig:catalogue_spatial_comparison_x_y}
\end{figure*}

\subsection{Cross-Matching}

To assess the reliability of our YSO catalogue, we have cross-matched both our initial YSO candidate sample and the final YSO association member sample with several published YSO and young-star catalogues. Because the reference samples differ in their input data, selection criteria, and age coverage, this comparison primarily serves to evaluate the consistency between independent catalogues. The cross-matching was performed using either Gaia DR3 source ID matching or a 1~arcsec positional match, depending on the reference catalogue. The results are summarised in Table~\ref{tab:appendix_crossmatch}. 

\subsubsection{Gaia DR3 Variable YSOs}

The Gaia DR3 variable YSO sample was constructed from Gaia multi-epoch photometry through the variability classification pipeline \citep{Rimoldini2023} and subsequently validated by \citet{Marton2023}. This catalogue contains 79\,375 YSO candidates that exhibited detectable optical variability during the 34-month Gaia observing window. Cross-matching by Gaia DR3 source ID yields 4812 matches with our initial candidate sample (0.65\%) and 646 matches with our final member sample (11.31\%). Because the sample construction paths are completely different, this limited overlap should be understood from both directions. From the perspective of \citet{Marton2023}, only 6.1\% (4812 out of 79\,375) of their variable YSO candidates appear in our initial sample, primarily because our sample is restricted to the $\sim$33 million Gaia sources with RV measurements, and our CMD selection and parallax precision cut ($\varpi/\sigma_{\varpi} \geq 5$) further exclude some of those sources. From our side, the match fraction rises from 0.65\% in the initial candidates to 11.31\% in the final members, indicating that the abundance of variable YSOs is significantly higher among the clustered association members, consistent with the expectation that association members are younger and more photometrically active.

\subsubsection{KYSO Catalogue}

The KYSO (Konkoly Optical YSO) catalogue \citep{Marton2023} is a literature compilation of 11\,671 optically identified young stars assembled from the Handbook of Star Forming Regions and subsequent published YSO catalogues, most of which have spectroscopic confirmation. The catalogue is biased towards well-studied star-forming regions and is not a uniform all-sky survey. We find 1429 matches in our initial candidate sample (0.19\%) and 348 matches in our final member sample (6.09\%). The low fraction of 0.19\% is largely a denominator effect given the large size of our initial candidate sample. From the KYSO viewpoint, only 12.2\% (1429 out of 11\,671) of KYSO sources appear in our initial candidates. \citet{Marton2023} note that only about 45\% of the KYSO sources entered the Gaia variability supervised classification and only about 40\% were included in the final Gaia DR3 variable YSO candidate catalogue, with roughly a quarter removed by Gaia data-quality cuts. Our sample also relies on Gaia data and additionally requires radial velocity measurements together with a location inside the PMS/YSO selection region on the CMD, which further limit the chance for a KYSO source to enter our sample. Among the KYSO sources that do enter our initial candidates, the survival rate into the final member sample is 24.4\% (348 out of 1429), the highest among the four reference catalogues. This high survival rate is consistent with the nature of KYSO: its sources mostly have spectroscopic confirmation and represent the highest-purity optical YSO sample, making them more likely to reside in real associations that pass the 6D HDBSCAN clustering and the isochrone test.

\subsubsection{Zari et al. (2018)}

The PMS catalogue of \citet{zari2018} was constructed from Gaia DR2 sources within 500~pc using photometric and astrometric criteria similar to our CMD-based selection. Cross-matching by 1~arcsec positional match yields 15\,780 matches in our initial candidate sample (2.13\%) and 1597 matches in our final member sample (27.95\%). This comparison is particularly relevant because both studies employ CMD-based PMS selection. The 27.95\% match fraction among final members is notable, considering that the sample of \citet{zari2018} is limited to 500~pc while our associations extend to $\sim$1~kpc and that \citet{zari2018} did not require radial velocities. The unmatched part mainly reflects these differences. The survival rate for this sample is 10.1\% (1597 out of 15\,780), the lowest among the four catalogues. The main reason is that \citet{zari2018} applied a 2D tangential velocity selection rather than 6D phase-space clustering, so their PMS sample contains more field contaminants and isolated sources that fail to form significant density enhancements in the six-dimensional space.

\subsubsection{SPYGLASS IV}

The SPYGLASS IV catalogue of \citet{Kerr2023} also uses Gaia DR3 data and HDBSCAN clustering to identify 116 young stellar associations within 1~kpc. Cross-matching by Gaia DR3 source ID yields 19\,854 matches in our initial candidate sample (2.68\%) and 2646 matches in our final member sample (46.32\%). The substantial overlap, where nearly half of our final member stars are also found in SPYGLASS IV, indicates that many of our identified associations lie within known young stellar structures in Gaia DR3. The remaining differences are mainly attributable to the wider age range of SPYGLASS IV (up to $\sim$50~Myr), differences in HDBSCAN parameters and clustering strategy, and our requirement for RV measurements.

\subsubsection{Summary of Cross-Matching}

Overall, the above cross-matching and survival-rate analysis support the reliability of our final YSO association catalogue. These results also confirm that our catalogue should be regarded as an optically selected YSO candidate sample with Gaia DR3 radial velocities, rather than a complete census of all YSO evolutionary stages. The catalogue is jointly shaped by the availability of Gaia RV measurements, optical detectability, and the CMD-based selection criteria adopted in this work.

\subsection{Spatial Comparison}

\subsubsection{Zari et al. (2018)}

Fig.~\ref{fig:catalogue_spatial_comparison_x_y} presents the spatial comparison between our YSO association catalogue and four reference samples in the Galactic $X$--$Y$ plane. 
Panel (a) compares our catalogue with the PMS sample of \citet{zari2018}. This comparison carries direct methodological significance because both studies adopt a CMD-based PMS selection. Within 500~pc, the high-density regions of the \citet{zari2018} PMS stars coincide almost completely with the distribution of our YSO associations, with major star-forming regions such as Taurus, Perseus, Orion, Sco-Cen, and Ophiuchus all showing strong spatial overlap.

\subsubsection{SPYGLASS IV}

Panel (b) compares our catalogue with the SPYGLASS IV young associations of \citet{Kerr2023}. In the nearby region that includes Orion, Sco-Cen, Vela, Taurus, Perseus, and CaMaS, the member star density distribution of \citet{Kerr2023} agrees very well with our YSO associations. The main differences concern more distant groups and the Cep-Her region. Several SPYGLASS IV groups at larger distances, such as NAS, Cone, Cepheus OB2/3, Vela-C, and NGC~2169, lie beyond the effective reach of our Gaia RV-selected sample and therefore lack corresponding YSO associations in our catalogue. In the region between the Radcliffe Wave and the Split, \citet{Kerr2023} identified the Cep-Her complex at an age of about 30~Myr, which appears as an extended structure in their density map. Our catalogue shows relatively few YSO groups in this area, consistent with the younger age limit of our sample, which is defined by isochrones no older than 20~Myr. This absence of older groups confirms the young nature of our catalogue.

\subsubsection{Kounkel et al. (2020)}

Panel (c) compares our catalogue with the young structures of \citet{kounkel2020a}, who used HDBSCAN on Gaia DR2 data to identify 8292 stellar groups within 3~kpc. We select from their catalogue only structures younger than 20~Myr and closer than 1~kpc, and construct density contours from the positions of the structure centres rather than from individual member stars. The two distributions agree very well over most of the common area, with Orion, Sco-Cen, Vela, Taurus, and Perseus all confirmed by both catalogues. Differences mainly arise for some \citet{kounkel2020a} groups at distances beyond about 700~pc, where our catalogue does not recover corresponding YSO associations owing to the Gaia RV distance limitation discussed above.

\subsubsection{Quintana et al. (2026)}

Panel (d) compares our catalogue with the OB association members of \citet{Quintana2026}, who applied HDBSCAN to a complete sample of $\sim$25\,000 O- and B-type stars and identified 56 OB associations within 1~kpc. In nearby regions, the agreement with our YSO associations is excellent, with Sco-Cen, Orion, Vela, Taurus, Perseus, and Lacerta all exhibiting strong overlap, confirming that both tracers follow the same star-forming complexes. Most of the more distant OB associations, such as Aur~OB1, Gem~OB3, Ara~OB1, and Car~OB4, lack corresponding YSO groups in our catalogue, again reflecting the distance limitation imposed by the Gaia RV requirement. Two cases highlight the complementarity of OB and YSO tracers. In the vicinity of Aquila~Rift, our catalogue and other young-star surveys detect clustered YSOs while no OB association is found, which indicates that this is a low-mass star-forming region lacking massive OB stars. Toward Sct~OB4 and Cyg~OB10, OB associations are present yet neither our catalogue nor the other reference samples recover corresponding young stellar groups. These directions lie along the Galactic plane where extinction is higher, and it is also possible that some of these OB associations are too evolved to still host the young YSOs traced by our sample.

\subsubsection{Summary of Spatial Comparison}

Overall, the spatial comparisons show that our YSO associations are distributed in the same regions as known young stellar populations. The agreement is particularly strong for nearby PMS-rich regions and for the large-scale structures traced by Gaia-based young association catalogues. The differences between our catalogue and the reference samples can be consistently understood through differences in stellar tracers, age limits, clustering strategies, the Gaia RV requirement, and optical visibility. These comparisons support the reliability of our final YSO association catalogue while also illustrating its main observational limitations.


\bsp	
\label{lastpage}
\end{document}